\documentclass[12pt]{iopart}

\usepackage[colorlinks=true,linkcolor=blue,urlcolor=blue,citecolor=blue,pdfusetitle]{hyperref}
\usepackage{color}
\usepackage[utf8]{inputenc}
\usepackage[usenames,dvipsnames]{xcolor}

\usepackage{amssymb}
\usepackage{graphicx}
\graphicspath{{graphics/}}
\usepackage{epsfig}
\usepackage{dcolumn}
\usepackage{bm}
\usepackage{mathrsfs}
\usepackage{multirow}
\usepackage[all]{xy}
\usepackage{pbox}
\usepackage{verbatim}
\usepackage{braket}
\usepackage{bbold}
\usepackage{xr}
\usepackage{float}
\usepackage{stackrel}

\newcommand{\eqref}[1]{(\ref{#1})}

\newcommand{\bc}[1]{{\color{blue}#1}}

\begin{document}
\raggedbottom

\title[]{Thermodynamics of coupled time crystals with an application to energy storage}

\author{Paulo J. Paulino$^{1}$, Albert Cabot$^{1}$, Gabriele De Chiara$^{2}$, Mauro Antezza$^{3,4}$, Igor Lesanovsky$^{1,5}$ and  Federico Carollo$^6$}

\address{$^1$Institut f\"{u}r Theoretische Physik and Center for Integrated Quantum Science and Technology, Universit\"{a}t T\"{u}bingen, Auf der Morgenstelle 14, 72076 T\"{u}bingen, Germany}

\address{$^2$ Centre for Quantum Materials and Technology, School of Mathematics and Physics, Queen’s University Belfast, Belfast BT7 1NN, United Kingdom}

\address{$^3$ Laboratoire Charles Coulomb (L2C) UMR 5221 CNRS-Universit\'e de Montpellier, F- 34095 Montpellier, France}

\address{$^4$ Institut Universitaire de France, 1 rue Descartes, F-75231 Paris Cedex 05, France}

\address{$^5$ School of Physics and Astronomy and Centre for the Mathematics and Theoretical Physics of Quantum Non-Equilibrium Systems, The University of Nottingham, Nottingham, NG7 2RD, United Kingdom}

\address{$^6$ Centre for Fluid and Complex Systems, Coventry University, Coventry, CV1 2TT, United Kingdom}


\date{\today}

\begin{abstract}
Open many-body quantum systems can exhibit intriguing nonequilibrium phases of matter, such as time crystals. In these phases, the state of the system spontaneously breaks  the time-translation symmetry of the dynamical generator, which typically manifests through persistent oscillations of an order parameter.
A paradigmatic model displaying such a  symmetry breaking  is the {\it boundary time crystal}, which has been extensively analyzed experimentally and theoretically. 
Despite the broad interest in these nonequilibrium phases, their thermodynamics and their fluctuating behavior remain largely unexplored, in particular for the case of coupled time crystals.
In this work, we consider two interacting  boundary time crystals and derive a consistent interpretation of their thermodynamic behavior. We fully  characterize their average dynamics and the behavior of their quantum fluctuations, which allows us to demonstrate the presence of quantum and classical correlations in both the stationary and the time-crystal phases displayed by the system.
We furthermore exploit our theoretical derivation to explore possible applications of time crystals as quantum batteries, demonstrating their ability to efficiently store energy.

\end{abstract}

\maketitle

\section{Introduction} 

Interacting open many-body quantum systems can display phases that break the time-translation symmetry of the dynamical generator, referred to as time crystals~\cite{PhysRevLett.109.160401,PhysRevLett.117.090402,PhysRevLett.121.035301,Gong2018}. 
Such a symmetry breaking constitutes a genuine nonequilibrium phenomenon~\cite{PhysRevLett.114.251603,PhysRevLett.110.118901}, that can emerge  due to the interplay  between driving, dissipation and interactions \cite{PhysRevLett.121.035301,Gong2018}.
In these settings, observables of the system, e.g., the average magnetization, exhibit persistent oscillations rather than approaching stationary values~\cite{HJCarmichael_1980, PhysRevLett.121.035301,PhysRevA.105.L040202,riera2020time}. 
These dissipative time crystals have been extensively explored theoretically and have been found to emerge in a variety of different settings \cite{Gong2018,Tucker2018,Buca2019,PhysRevLett.122.015701,Lazarides2020,Seibold2020,PhysRevB.100.054303,PhysRevA.108.062216,PhysRevB.103.184308,PhysRevLett.121.035301,PhysRevA.108.L041303,Krishna2023,PRXQuantum.5.030325,PhysRevA.108.023516}. Several experimental observations on the realization of time crystals have also been reported~\cite{Kessler2019,doi:10.1126/science.abo3382,Wu2024dissipative,Kongkhambut_2024,PhysRevA.109.063317}.

\begin{figure}[t!]
    \centering
     \includegraphics[width=\linewidth,height=\textheight,keepaspectratio]{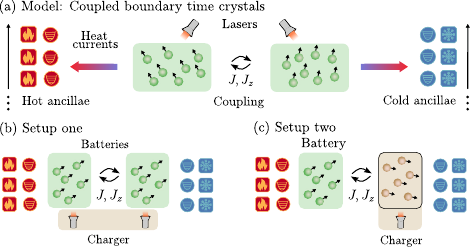}
    \caption{\textbf{ Collision-model setup for a system of two interacting boundary time crystals.} (a) A hot ancillary system, in a thermal state with inverse temperature $\beta_1$, interacts with one boundary time crystal (BTC), while a cold ancillary system, in a thermal state with inverse temperature $\beta_2$, interacts with the other BTC. The two BTCs are also driven by two independent lasers and coupled through an interaction Hamiltonian parametrized by the coupling strengths $J$,  in the $x$ and $y$ directions, and $J_z$, in the $z$ direction (see main text). (b) Two interacting BTCs working as batteries driven by the lasers. (c) One BTC charging the other one via seeding crystallization in time.}
    \label{fig:Fig1}
\end{figure}

Technological applications of time crystals have been proposed in the context of quantum metrology~\cite{Montenegro2023,PhysRevLett.132.050801, PhysRevA.109.L050203,yousefjani2024discrete} and of quantum thermodynamics, for instance in nonequilibrium quantum engines~\cite{PhysRevA.108.023516, PhysRevLett.125.240602}.
In order to be able to assess this potential, there is the  need of a consistent thermodynamic description that accounts for their intrinsic nonequilibrium nature~\cite{levy2014, deffner2019quantum}.
One way to achieve this is by exploiting a collision-model approach for modelling their interaction with the surroundings \cite{carollo2023quantum, cusumano2022quantum, de2018reconciliation, PhysRevLett.126.130403,PhysRevA.96.032107,ciccarello2022quantum}.
Within this framework, the  environment consists of a set of ancillae that individually, and unitarily, interact with the system [see Fig.~\ref{fig:Fig1}\bc{(a)}]. 
By evaluating the energy and the information exchange with the ancillae, it is possible to provide a fully consistent characterization of the thermodynamic quantities related to the system~\cite{de2018reconciliation,PhysRevX.7.021003,cusumano2022quantum,PhysRevResearch.4.023230,PhysRevResearch.3.013165}. 

Quantum thermodynamics is also concerned with the use of quantum systems for energy storage. Quantum devices performing this task are also known  as quantum batteries~\cite{PhysRevE.87.042123,PhysRevLett.128.140501,PhysRevLett.132.210402,campaioli2023colloquium,PhysRevE.106.054119,PhysRevA.97.022106,PhysRevB.99.205437} and are being studied in the context of closed~\cite{PhysRevE.100.032107,PhysRevB.100.115142,PhysRevE.103.042118} and open quantum systems~\cite{Mitchison2021chargingquantum, 10.1063/1.5096772,PhysRevB.99.035421, PhysRevLett.122.210601,PhysRevResearch.2.033413,PhysRevA.108.L050201, PhysRevA.107.042419,Chang_2021,PRXQuantum.5.030319}, for which experimental realisations have been reported~\cite{Hu_2022,PhysRevLett.131.260401,batteries8050043,PhysRevA.106.042601}.
An important platform in this field are driven-dissipative quantum batteries, where a persistent external energy input is used to compete with dissipation effects in order for the system to store a finite amount of energy~\cite{PhysRevLett.122.210601,PhysRevResearch.2.033413,PhysRevApplied.20.044073,PhysRevLett.132.210402,Beleño_2024}. However, the possibility of exploiting time crystals as quantum batteries in these driven-dissipative settings  has remained unexplored so far. 

In this work, we consider a system consisting of two interacting boundary time crystals and explore its potential for energy storage [see Fig.~\ref{fig:Fig1}].  
We investigate the coupled systems focusing on two different setups. 
In one setup, both boundary time crystals are used for energy storage and the charging occurs due to input from external lasers [see Fig.~\ref{fig:Fig1}\bc{(b)}].
This setup is also known as parallel charging process~\cite{PhysRevLett.120.117702,PhysRevE.102.052109}.
In the other setup, one boundary time crystal is driven by a laser and is considered to be the charger, while the other one is the battery and is not subject to any external driving.
In this way, energy can be transferred from the charger to the battery, which in the time-crystal phase happens  via seeding crystallization in time \cite{PhysRevLett.128.080603} [see Fig.~\ref{fig:Fig1}\bc{(c)}]. 
Through a collision-model approach, we characterize the thermodynamics of these systems which allows us to define a measure of efficiency for the charging process, given by the ratio between stored energy  and charging cost.
We then compare the energy stored in the batteries and their  efficiency when operating in a variety of time-crystal and stationary phases emerging in both setups.
Our analysis shows that when the system is found in the time-crystal phase it can store more energy, with a higher  efficiency compared to the stationary phase. 
We also analyze quantum fluctuations in the system and  compute quantum and classical correlations between the coupled boundary time crystals.
We find that the first setup considered [Fig.~\ref{fig:Fig1}\bc{(b)}] exhibits entanglement in the stationary phase, while the second one [Fig.~\ref{fig:Fig1}\bc{(c)}] displays entanglement within the time-crystal phase.

\section{Model}

We consider two ensembles of atoms with $N$ two-level atoms each. Their bare Hamiltonian is ($\hbar =1$) ${H^{{(j)}}_{\rm at} = \omega_{\rm at}S_z^{(j)}/\sqrt{2}}$, where $j=1,2$ refers to the atomic systems, and $\omega_{\rm at}>0$ is the energy splitting between the  atomic levels.
Here, ${S_\alpha^{(j)} = 1/\sqrt{2}\sum_{m=1}^{N} \sigma_{\alpha,m}^{(j)}}$ is a collective operator, representing the atoms in the different ensembles, such that $\sigma^{(j)}_{\alpha, m}$ is the Pauli matrix in the $\alpha$ direction, $\alpha= x,y,z$, for the $m$-th atom in the $j$-th ensemble.
The two ensembles interact collectively through the Hamiltonian
\begin{equation}
    H_{\rm int} = \frac{J}{N}\left( S_x^{(1)}S_x^{(2)} + S_y^{(1)}S_y^{(2)}\right) + \frac{J_{z}}{N} S_z^{(1)}S_z^{(2)} ,
    \label{int_Hamiltonian}
\end{equation}
where $J$ is the coupling strength in both $x$ and $y$ direction, while $J_z$ is the one in the $z$ directions.
The factor $1/N$ is included to ensure a well defined thermodynamic limit~\cite{PhysRevA.8.2517}. The two ensembles are driven by external lasers that are modelled by the Hamiltonian
\begin{equation}
    H_{\rm las}^{(j)} = \frac{\Omega_j}{\sqrt{2}}(S_-^{(j)}e^{i\omega_{\rm las}t} +S_+^{(j)}e^{-i\omega_{\rm las}t}) ,
\end{equation}
where ${S_{\pm}^{(j)}=S_x^{(j)} \pm i S_y^{(j)}}$, ${\omega_{\rm las}}$ is the laser frequency, which we assume  to be the same for both ensembles, and $\Omega_{j}$ is the Rabi frequency. 
The Hamiltonian of the full system is $H_{\rm s} = H^{{(1)}}_{\rm at} + H^{{(2)}}_{\rm at} + H^{{(1)}}_{\rm las} + H^{{(2)}}_{\rm las} + H_{\rm int}$, which in the frame rotating with the laser frequency, becomes
\begin{equation}\label{eq:Hamiltonian}
    H_{\rm s}^{\rm rot}= \frac{\Omega_{1}}{\sqrt{2}} S_x^{(1)} +  \frac{\Omega_{2}}{\sqrt{2}} S_x^{(2)} + \frac{\delta}{\sqrt{2}} S_z^{(1)} +  \frac{\delta}{\sqrt{2}} S_z^{(2)} + H_{ \rm int} .
\end{equation}
Here $\delta=\omega_{\rm at} - \omega_{\rm las}$ is the detuning between the laser frequency and the energy splitting $\omega_{\rm at}$. 
Each of the atomic ensembles is coupled to a thermal bath with inverse temperature $\beta_j$. The driven-dissipative dynamics is then governed by the master equation $\dot{\rho} = \mathcal{L}[\rho]$, where 
\begin{equation}\label{eq:Lindblad}
    \mathcal{L}[\rho] = -i[H_{\rm s}^{\rm rot}, \rho] +\sum\limits_{j,\alpha} \left[ L_\alpha^{(j)} \rho (L_\alpha^{(j)})^\dag - \frac{1}{2}\left\{(L_\alpha^{(j)})^\dag L_\alpha^{(j)}, \rho \right\} \right] .
\end{equation}
The jump operators that describe the dissipative processes are
\begin{equation}\label{eq:JumpOperators}
L_+^{(j)} = \sqrt{\frac{\kappa n_j}{N}} S_+^{(j)} ,L_-^{(j)} = \sqrt{\frac{\kappa (1+n_j)}{N}} S_-^{(j)} ,
\end{equation}
where $n_j=(e^{{\beta_j \nu}} - 1)^{-1}$ is the occupation number of the environment degrees of freedom,  $\beta_j$ is the inverse temperature  and $\nu$ the energy scale for the environment.
Here, the parameter $\kappa$ encodes the decay rate. 
Essentially, the considered system consists of two interacting boundary time crystals 
and immersed in two separate  environments at finite temperature [see sketch in Fig.~\ref{fig:Fig1}\bc{(a)}].

\section{Quantum thermodynamics}

The master equation describing the dynamics of the  system, Eq.~\eqref{eq:Lindblad}, is a so-called local master equation, characterized by jump operators which do not implement transitions between the eigenstates of the system Hamiltonian~\cite{de2018reconciliation,breuer2002theory,PhysRevResearch.4.023230,PhysRevResearch.3.013165}.
This equation could be obtained by considering that the interaction between the subsystems and the laser driving do not affect the coupling between the bare systems and their environment~\cite{HJCarmichael_1973,levy2014,Hofer_2017}.    
Although this phenomenological approach is successfully employed to model and interpret several experiments~\cite{brennecke2007,brennecke2008,purdy2010,PhysRevLett.127.043602,RevModPhys.85.553}, it can lead to thermodynamic inconsistencies if the relevant  quantities are not properly defined.
For example, it can lead  to nonequilibrium stationary states featuring persistent heat currents that do not obey  Spohn’s inequality~\cite{10.1063/1.523789,levy2014,alicki1979,PhysRevA.108.023516}.
In other words, these master equations may apparently violate the laws of thermodynamics as formulated for a system weakly coupled to an external thermal bath~\cite{deffner2019quantum}. 
Nevertheless, a consistent description of systems governed by local master equations can be obtained, for instance, by considering a collision-model interpretation of their interaction with the surroundings \cite{PhysRevA.96.032107, ciccarello2022quantum, PhysRevLett.126.130403, cusumano2022quantum, de2018reconciliation}. 

In our setup, the two atomic ensembles are coupled to different  thermal baths.
Within the collision-model approach~\cite{PhysRevA.96.032107, ciccarello2022quantum, PhysRevLett.126.130403, cusumano2022quantum, de2018reconciliation}, we model each environment as a stream of ancillary units consisting of bosonic modes.
The dynamics follows from the joint unitary time evolution of the system and the ancillae.
Each ancilla interacts with the system only once and for a time $\delta t$, after which it can be discarded from the description. 
The process is then repeated with the next ancilla. In the continuum limit  $\delta t\rightarrow 0$, the sequence of collisions can reproduce the dynamics of the master equation in Eq.~\eqref{eq:Lindblad} [see Fig.~\ref{fig:Fig1}\bc{(a)}].
The ancillary systems are described by the bare Hamiltonian $H_{\rm e} = {\sum_k}[ H_{{\rm e}_k}^{{(1)}} + H_{\mathrm{e}_k}^{{(2)}}]$, where $H_{\mathrm{e}_k}^{{(j)}}= \nu a^{{(j)\dag}}_k a_k^{(j)}$ is the free Hamiltonian of the ancilla  corresponding to the $k$-th collision in system $j$, $\nu $ is its bare frequency, and $a^{{(j)}}_k$ $ {(a_k^{{(j)\dag}})}$ are the bosonic annihilation (creation) operators.
The Hamiltonian describing the $k$-th ancilla interacting with the $j$-th atomic ensemble is
\begin{equation}\label{eq:colision_interaction}
    H_{{\rm se}_k}^{(j)}(t) = \sqrt{\frac{\kappa}{N \delta t}} \theta(t, k)\left(a_k^{(j)\dag} S_-^{(j)} + a_k^{(j)} S_+^{(j)} \right) ,
\end{equation}
where $\theta(t, k)$ is defined as $\theta(t, k) = 1$ for $k\delta t < t < (k+1)\delta t$ and $0$ otherwise.
In order to ensure a completely positive  Markovian  dynamics for the system, we consider the \textit{ancillae} to be initially in a product state, with each ancilla being in the thermal state with occupation number given by $\langle (a_k^{(j)})^\dag a_k^{(j)} \rangle = n_j$ [see definition after Eq.~\eqref{eq:JumpOperators}].
This collision-model description of the system-environment interaction allows us to  derive a consistent thermodynamic description of the system~\cite{de2018reconciliation,carollo2023quantum}.

To this end, we consider the thermodynamic behaviour of the two atomic ensembles in the laboratory frame, in which the contribution from the bare atomic frequency, $H_{\rm at}$, is taken into account. 
Thus, we start with $H_\mathrm{s}$ and we consider the variation of the density matrix after a collision, 
\begin{equation}
    \Delta \rho_{\mathrm{se}_k} = V_k \left(\rho_\mathrm{s} \otimes \rho_{\mathrm{e}_k} \right)V_k^\dag - \rho_{\mathrm{s}}\otimes \rho_{\mathrm{e}_k} ,  
\end{equation}
where  $\rho_{{\rm  e}_k} = \rho_{{\rm e}_k}^{(1)}\otimes \rho_{{\rm e}_k}^{(2)}$. 
Here, the unitary evolution of the system and the ancillae reads $V_k = \mathcal{T}e^{-i\int_t^{t+\delta t}\left( H_{\rm s} + H_{{\rm e}_k}^{(1)}+ H_{{\rm  e}_k}^{(2)} +  H_{{\rm  se}_k}^{(1)} +  H_{{\rm se}_k}^{(2)}\right){\rm d}t^\prime}$. 
Also, $\mathcal{T}$ is a time-ordering operator,  which is needed given that in the laboratory frame $H_{\rm s}$ is time-dependent due to the laser driving and given that the  interaction with the ancillae is also time-dependent.
The heat dissipated by the $j$-th system after a collision is defined as minus the energy variation of the ancilla,
\begin{equation}
    \Delta{Q}^{(j)}(t) =  -\Tr_k\{H_{{\rm e}_k}^{(j)}\Delta\rho_{{\rm se}_k} \} ,
\end{equation}
where $t=k \delta t$.
Similarly, we define the variation of the internal energy of the system after a collision as  
\begin{equation}
\Delta U = \Delta U_{\rm s}+ \Delta W_{\rm las} ,
\end{equation}
where $\Delta W_{\rm las} = \Tr_k\left\{ \Delta H_{\rm las} \rho_{{\rm se}_k} \right\}$ is the work done on the system due to the time-dependence of the laser Hamiltonian and $\Delta U_{\rm s} = \Tr_k\{H_{\rm s} \Delta\rho_{{\rm se}_k}\}$. 
We find the first law of thermodynamics by considering that the total work input in the system can be computed as the difference between the variation of the internal energy of the system  and the total dissipated heat
\begin{equation}\label{eq:Work_first_law}
\dot{W} = \lim_{\delta t \rightarrow 0}\frac{1}{\delta t} \left( \Delta W_{\rm env} - \Delta W_{\rm las}\right) = \dot{U}_{\rm s} - \dot{Q} ,
\end{equation}
where $\dot{A} = \lim_{\delta t \rightarrow 0}\Delta A/\delta t$, $\dot{Q}=\dot{Q}^{(1)}+\dot{Q}^{(2)}$ and $\Delta W_{\rm env}$ is the work done by turning on the interaction between the system and the different  ancillae (see \ref{ap:thermodynamics} for  details on the second law of thermodynamics for the coupled boundary time crystals).  
In the following, we exploit the above thermodynamic quantities to characterize energy storage and its efficiency in two different setups of the coupled boundary time crystals [see an illustration in Fig.~\ref{fig:Fig1}\bc{(b)-(c)}]. 

\section{Energy storage with interacting time-crystal batteries}\label{sec:setup1}

In this section we consider the setup in which the boundary time crystals  act as a battery charged by the external lasers and the interaction with the environment [see Fig.~\ref{fig:Fig1}\bc{(b)}], such that the system follows a parallel charging process~\cite{PhysRevLett.120.117702,PhysRevE.102.052109}. 
To lay the foundation for this investigation we first characterize the phase diagram of the system in the thermodynamic limit.
We consider both atomic systems to have the same parameters, $\Omega_1=\Omega_2=\Omega$ and $\delta =0$, with all atoms initially in the ground state of the bare atomic Hamiltonian. 

We define the average expected values ${m_{\alpha}^{(j)} = \langle S_{\alpha}^{{(i)}}/N \rangle}$, whose equations of motion can be obtained from the master equation in  Eq.~\eqref{eq:Lindblad}.
In the thermodynamic limit, $N\rightarrow \infty$, the average dynamics of the system is exactly described by the mean-field equations~\cite{carollo2021,carollo2024applicabilitymeanfieldtheorytimedependent}, which  in the rotating frame read
\begin{eqnarray}\label{eq:Symmetric_equations}
    \dot{m}_{x} &= (J-J_z)\sqrt{2}m_{y}m_{z}  + \kappa \sqrt{2} m_{x}m_{z} ,\\
    \dot{m}_{y} &= (J_z-J)\sqrt{2}m_{x}m_{z}   +\kappa \sqrt{2}m_{y}m_{z} - \Omega m_{z},\\
    \dot{m}_{z} &= - \kappa\sqrt{2}(m_{x}^2 + m_{y}^2) + \Omega m_{y}.
\end{eqnarray}
Here $m_\alpha=m_\alpha^{(1)}=m_\alpha^{(2)}$, as for the considered initial conditions both systems feature the same mean-field dynamics. 
We note that the mean-field equations are not dependent on the temperature of the environments~\cite{carollo2023quantum}.
Besides the conservation of the norm of the vector $\vec{m} = [m_x(t),m_y(t),m_z(t)]$, the system of equations features another conserved quantity. This is given by the expression~\cite{PhysRevLett.121.035301}
\begin{equation}\label{eq:ConvervedQuantity}
        \Gamma = i\left[ \lambda_- \log\left( \lambda_+ \eta_+  -  \frac{\Omega}{\sqrt{2}} \right)  -  \lambda_+ \log\left( \lambda_- \eta_-  -  \frac{\Omega}{\sqrt{2}} \right)\right] ,
\end{equation}
where $\lambda_\pm  = \sqrt{2}(\kappa \pm i(J-J_z))$, and $\eta_\pm = (\pm i m_x + m_y)/\sqrt{2}$, so that its value depends both on system parameters and initial conditions.
The system features a stationary phase, a time-crystal phase, and a region in which both phases coexist. In the stationary phase, we find the following stationary solutions for the magnetizations
\begin{eqnarray}\label{eq:Stationary_sols}
    m_x^\mathrm{s} &= \frac{1}{\sqrt{2}}\frac{\Omega (J_z-J)}{(J_z-J)^2 + \kappa^2}  ,\\
    m_y^\mathrm{s} &= \frac{1}{\sqrt{2}}\frac{\kappa \Omega}{(J_z-J)^2 + \kappa^2},\\
    m_z^\mathrm{s} &=  \frac{\pm1}{\sqrt{2}}\sqrt{1 - \frac{\Omega^2}{(J_z-J)^2 + \kappa^2}},
\end{eqnarray}
which exist as long as the condition $(J_z-J)^2 + \kappa^2 \ge \Omega^2$ is satisfied.
We note that the magnetization in the $z$-direction has two possible values, a negative one which is stable and a positive one which is unstable.
When the above condition is not fulfilled, the system displays oscillations. 
The time-crystal phase is characterized by a continuous family of oscillatory solutions, or limit cycles, each one associated with a possible value of the conserved quantity in Eq.~\eqref{eq:ConvervedQuantity}.
This is similar to what is observed in the single boundary time-crystal case \cite{HJCarmichael_1980,DRUMMOND1978160, PhysRevLett.121.035301}. 
These different oscillatory solutions can be accessed by varying the initial conditions.

Concerning thermodynamic quantities, the heat exchanged with the $j$-th environment is given by 
\begin{equation}
    \dot{Q}^{(j)}  = - \frac{\kappa\nu }{N} \left< (S_x^{(j)})^2   +   (S_y^{(j)})^2   +  \sqrt{2}(2n_j  + 1 ) S_z^{(j)} \right>,
\end{equation}
(see \ref{ap:thermodynamics} for  details).
The heat currents are characterized by a dominant part, proportional to $N$, and another contribution dependent on the temperature (through $n_j$) that is intensive with $N$. This results from the factorization $\langle S_{\alpha}^{{(i)}}S_{\beta}^{{(j)}}/N^2 \rangle = m_{\alpha}^{{(i)}}m_{\beta}^{{(j)}} + \mathcal{O}(N^{-1})$, where $\mathcal{O}(N^{-1})$ denotes a contribution of order $N^{-1}$, which is exact in the thermodynamic limit~\cite{carollo2023quantum}. Therefore, in this limit, the heat exchanged (per atom) with the $j$-th  environment becomes 
\begin{equation}
    \dot{q}^{(j)} = -\kappa \nu \left[ (m_x^{(j)})^2 + (m_y^{(j)})^2\right],
\end{equation}
where we denote $\lim_{N\rightarrow \infty}\dot{Q}^{(j)}/N = \dot{q}^{(j)}$. 
Noteworthily,  the heat $\dot{q}^{(j)}$ does not depend on the temperature of environment since the latter  does not appear in the mean-field dynamics~\cite{carollo2023quantum}. 
Also, the heat is always negative, meaning that energy is dissipated and  always flows from the system to the environment.
This suggests that no efficient heat engine could be devised from our model~\cite{PhysRevE.106.014143}.
The time-average of $\dot{q}(t)$ during a long time-window, reduces the first law, Eq.~\eqref{eq:Work_first_law}, to 
\begin{equation}
    \bar{\dot{w}} = - \bar{\dot{q}} ,
\end{equation}
where $\bar{\dot{a}} = \lim_{t\rightarrow \infty} t^{-1} \int_0^t \dot{a}(t^\prime) \mathrm{d}t^\prime$.
In deriving this expression, we use the fact that the internal energy rescaled by $1/N$ does not grow indefinitely with time, such that $t^{-1}\int_0^t\dot{u}\mathrm{d}t^\prime= t^{-1}(u_t - u_0)$ goes to zero for large times $t$. 

\begin{figure}
    \centering
    \includegraphics[width=\linewidth,height=\textheight,keepaspectratio]{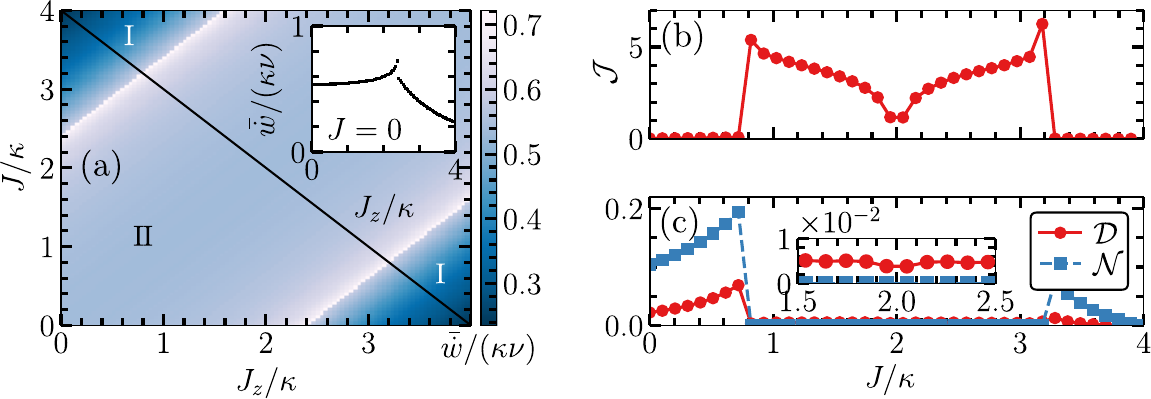}
    \caption{\textbf{Time-averaged work input and  correlations}. (a) Total time-averaged work input for varying $J_z$ and $J$.  Region I denotes the stationary phase, while region II the time-crystal phase. The inset shows the phase transition for $J=0$. The other parameters are $\delta = 0$, $ n_1 = n_2 = 0$ and $\Omega_1/2 = \Omega_2/2 = \kappa$. The initial state is $\vec{m}^{(j)} = [0, 0, -1/\sqrt{2}]$ for both systems in every point of the phase-diagram.
    (b) Time-averaged classical correlations $\mathcal{J}$ and (c) time-averaged quantum discord $\mathcal{D}$ and negativity $\mathcal{N}$ as function of $J$, such that $J+J_z=4\kappa$ [see the diagonal in panel (a)]. The mean-field quantities were evolved from $t=0$ to $t\kappa=10^3$ and the average was computed by integrating over the second half of the evolution time. The correlations (see Sec.~\ref{sec:QuantumFluctuations}) were computed considering trajectories between $t=0$ and $t\kappa=200$, and the integration was done over the second half of such interval. The remaining parameters are $\Omega_1=\Omega_2=2\kappa$, and $n_1=n_2=0$.}
    \label{fig:Thermodynamics}
\end{figure}

In Fig.~\ref{fig:Thermodynamics}\bc{(a)} we show the behavior of the time-averaged work input in the system, $\bar{\dot{w}}$, which exhibits two distinct dynamical regions.
In region I, the mean-field dynamics has a stationary solution, while in region II it only exhibits persistent oscillations.
However, in region I, we also observe oscillatory solutions, indicating a bistable regime with coexistence of oscillatory and stationary solutions approached by different initial conditions, see Fig.~\ref{fig:batteries}\bc{(a)} [see also   \ref{ap:meanfield}
 for more details].
For the stationary solutions, the total time-averaged work input per particle is given by ${\bar{\dot{w}} = \kappa \nu \Omega^2/[(J-J_z)^2 + \kappa^2]}$. 
This shows that the work input grows higher close to the phase transition, approaching  ${\bar{\dot{w}} = \kappa \nu}$ near the transition point.
In contrast, the work input goes to zero when $(J-J_z)^2 \gg \kappa^2$. 
In Fig.~\ref{fig:Thermodynamics}\bc{(a)}, we observe that the work input is higher in the time-crystal phase than in the stationary one, see inset in Fig.~\ref{fig:Thermodynamics}\bc{(a)}. 

\subsection{Quantum and classical correlations}\label{sec:QuantumFluctuations}

Beyond the mean-field description, it is possible to investigate the behavior of quantum fluctuations within the system~\cite{benatti2015dissipative,Goderis1990,carollo2023quantum,PhysRevA.108.062216,Benatti_2018}. Quantum fluctuations can be described through the operators
\begin{equation}
 F_\alpha^{(j)} = \frac{S_\alpha^{(j)} - \langle S_\alpha^{(j)}\rangle}{\sqrt{N}} ,
\end{equation}
which describe the deviation of  $S_\alpha^{(j)}$ from its average value. 
These operators account for the leading correction to
${\langle S_\alpha^{(j)} S_\beta^{(k)}\rangle /N^2 - m_\alpha^{(j)} m_\beta^{(k)} \approx \langle F_\alpha^{(j)} F_\beta^{(k)}\rangle/N}$ and capture both thermal and quantum effects.
In the thermodynamic limit, $N\to\infty$, the fluctuations behave as bosonic operators and they follow Gaussian statistics with zero average.
Therefore, the states of the fluctuations are fully characterized by the $6\times 6$ covariance matrix ${G^{(j,k)}_{\alpha\beta} = \langle\{F_\alpha^{(j)}, F_\beta^{(k)} \} \rangle/2}$~\cite{PhysRevA.79.052327,PhysRevA.106.012212}.
Additionally, these operators obey the commutation relations 
\begin{equation}\label{eq:fluctuations_comutation}
    \lim_{N\rightarrow \infty }    -i\langle[F_\alpha^{(j)}, F_\beta^{(k)}]\rangle =\sqrt{2}\sum_{\gamma}{\epsilon}_{\alpha\beta\gamma}\delta_{jk}m_\gamma^{(j)} = \delta_{jk} {s}_{\alpha\beta} ,
\end{equation}
where $\epsilon_{\alpha\beta\gamma}$ is a Levi-Civita tensor, $\delta_{jk}$ is the Kronecker delta and ${s_{\alpha\beta}}$ is a symplectic matrix. 
These commutation relations entail the fact that the fluctuations remain quantum in the thermodynamic limit and can thus be used to describe collective non-classical correlations~\cite{benatti2015dissipative,PhysRevA.106.012212}.
The dynamics of the covariance matrix can be  computed through  Eq.~\eqref{eq:Lindblad} in the Heisenberg picture, and it is given in    \ref{ap:fluctuations}.

To evaluate the correlations among the boundary time crystals, we can analyse the fluctuations in the reference frame which rotates solidly with the mean-field observables, such that the commutation relations in Eq.~\eqref{eq:fluctuations_comutation} become time independent~\cite{PhysRevA.105.L040202, PhysRevA.108.062216}.
To obtain this rotated frame, we first consider a time-dependent rotation matrix $R(t)$ that evolves the mean-field equations, ${\vec{m}(t) = R(t) \vec{m}(0)}$.
To find $R(t)$, we rewrite the mean-field equations as $\dot{\vec{m}}(t)=D(t) \vec{m}(t)$, where $D(t)$ is a time-dependent matrix and  $\vec{m}=[m_x^{(1)},m_y^{(1)},m_z^{(1)},m_x^{(2)},m_y^{(2)},m_z^{(2)}]$ is the vector of the mean-field variables.
The rotation $R(t)$ follows from the integration of $\dot{\vec{m}}(t)$ (see \ref{ap:fluctuations} for more details).
Thereafter, we define the rotated fluctuation operators $\vec{X} = R^T(t) \vec{F} = [x_1, p_1, w_1, x_2, p_2, w_2]$, where $\vec{F}=[F_x^{(1)},F_y^{(1)},F_z^{(1)},F_x^{(2)},F_y^{(2)},F_z^{(2)}]$ is the vector of the quantum fluctuation operators of the atomic ensemble. 
In the rotated reference frame, the quantum fluctuations obey a two-mode bosonic algebra $[x_j,p_k]=i\delta_{jk}$ and $w_{j}$ commutes with all the other operators.
Hence,  $w_{j}$ does not contribute to the dynamics of the bosonic fluctuations. 
Finally, we define the rotated covariance matrix  $\bar{G} = R^T(t) G(t) R(t)$, which represents a two-mode Gaussian system~\cite{benatti2015dissipative} (see   \ref{ap:fluctuations} for more details).

Using the symplectic eigenvalues of $\bar{G}$~\cite{PhysRevA.98.022335}, we can evaluate the von Neumann entropy, defined as $S(\rho) = -\Tr[\rho \ln \rho]$, for the coupled boundary time crystals~\cite{carollo2023quantum}.
Additionally, we can use the results on Gaussian bosonic systems to calculate classical and quantum correlations within the system, denoted by $\mathcal{J}$, and the quantum discord, $\mathcal{D}$, respectively.
Here, we note that the mutual information between the two modes is $I = \mathcal{D} + \mathcal{J}$.
In addition to that, we can also evaluate the negativity, $\mathcal{N}$, which measures collective entanglement between fluctuations of the coupled systems~\cite{PhysRevLett.105.020503, PhysRevLett.105.030501} (see   \ref{app:QuantumInformation} for more details).

The behavior of correlations is shown in Fig.~\ref{fig:Thermodynamics}\bc{(b)-(c)}.
We evaluate the quantities calculated from fluctuations along the diagonal line in Fig.~\ref{fig:Thermodynamics}\bc{(a)} given by $J+J_z=4\kappa$, such that we can visualize correlations in both phases. 
When starting from the time-crystal phase, the classical correlations exhibit a sudden jump from the stationary phase to the time-crystalline phase. 
On the other hand, the discord and the entanglement negativity have higher values in the stationary phase as compared  to the time-crystal one. 
We note that in the time-crystal phase both discord and classical correlations are minimized at the point $J=J_z$.
Remarkably, there is no entanglement in the time-crystal phase. 

When $J=J_z$ the mean-field equations recover those of a single boundary time crystal.
This can be understood in the context of synchronization blockade~\cite{PhysRevLett.118.243602,PhysRevA.108.022216,PhysRevA.97.013811}, where the systems dynamics are effectively decoupled in spite of the interaction terms. 
Also, along this line of the phase-diagram, an analytical solution for the mean-field equations is known~\cite{PhysRevA.105.L040202}, allowing us to write down the Hamiltonian and jump operators describing the dynamics of the Gaussian fluctuations~\cite{PhysRevA.105.L040202,PhysRevA.108.062216}. 
In particular, we find the Hamiltonian in the rotated frame to be 
\begin{equation}\label{eq:Hamiltonian_fluc_setup1}
    H_{\rm f} = J\left[x_1x_2 + p_1p_2 \right] ,
\end{equation}
which is time-independent and represents just an exchange of excitations term.
Also, each mode is affected by the jump operators 
\begin{equation}
V^{(j)}(t) = x_j - i\sqrt{2}m_z^{(j)}(t) p_j 
\end{equation}
and $[V^{(j)}(t)]^\dag$.  
This analysis shows that although the atomic systems become decoupled at the mean-field level when $J=J_z$, they still develop collective  correlations which are captured by quantum fluctuations.

\subsection{Energy storage}

In order to quantify the amount of energy stored  within the system, we consider the difference between the instantaneous intensive energy in the system and the energy contained at the initial time, $\mathcal{E}(t) = \Tr\{[\rho(t)-\rho(0)] H_{\rm ref}\}/N$,  where $H_{\rm ref}$ is a  Hamiltonian operator encoding the  accessible  energy which can be stored in the system~\cite{A.E.Allahverdyan_2004,carollo2021,PhysRevLett.122.047702}. In particular, we shall consider $H_{\rm ref}$ to be related to the bare energy of the atoms, which is also justified by the fact that typically $\omega_{\rm at} \gg J, J_z, \Omega$. The mean-field description of the system provides a way to estimate the intensive quantity  $\mathcal{E}(t)$   in the thermodynamic limit.

When the lasers are turned off, the dissipation brings the system toward the ground state of the bare atom Hamiltonian from which no energy can be extracted \cite{Sparaciari2017,Brown_2016}. 
To store energy within the system, one thus needs an external energy input from the laser which constitute a charging ``cost" necessary to sustain a nonequilibrium state featuring a finite energy. 
As a consequence, an important quantity to characterize driven-dissipative batteries is the charging efficiency, which measures the ratio between the stored energy and the total energy input to keep the battery charged~\cite{campaioli2023colloquium,PhysRevLett.122.210601,PhysRevResearch.2.033413}
\begin{equation}
    \eta(t) = \frac{\mathcal{E}(t)}{\int_{0}^t \dot{w}(u) \mathrm{d}u} = \frac{\mathcal{E}(t)}{\Delta u_\mathrm{s}(t)  - \int_0^t \dot{q}(u) \mathrm{d}u} ,
\end{equation}
where $\Delta u_\mathrm{s}(t) = \Tr\{[\rho(t)-\rho(0)] H_{\rm s}\}/N$, and the second equality follows from the first law of thermodynamics given in Eq.~\eqref{eq:Work_first_law}.
Here, $H_{\rm s}$ is the Hamiltonian of the system in the laboratory frame and we note that $\dot{q}$ is always negative. Given the assumption $\omega_{\rm at} \gg J, J_z, \Omega$, we can consider that $ H_\mathrm{s} \approx  H_{\rm at}^{(1)} + H_{\rm at}^{(2)} $ so that $\Delta u_\mathrm{s}(t)$ is well approximated by the variation of the bare atom energy in both atomic ensembles. 
The quantity $\eta(t)$ provides the efficiency of the charging process when $\mathcal{E}(t)>0$, otherwise the process is actually removing energy from the system.
Given that the stored energy  is bounded in time while the cost to keep the energy stored grows linearly with time, $\eta(t)$ approaches zero as the dynamics total time goes to infinity.

\begin{figure}[t]
    \centering
    \includegraphics[width=\linewidth,height=\textheight,keepaspectratio]{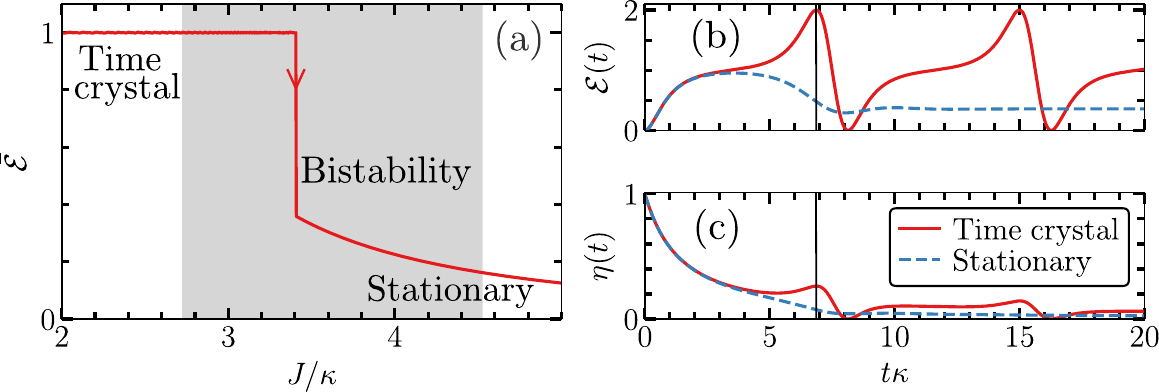}
    \caption{\textbf{Stored energy and charging efficiency}.  (a) Time-averaged stored energy in both atomic ensembles, defined as  $\bar{\mathcal{E}}= \lim_{t\rightarrow \infty} t^{-1} \int_0^t \mathcal{E}(t^\prime) {\rm d}t^\prime$, in units of $\omega_{\rm at}$.
    Here, we consider the initial condition to be with all atoms in the ground state. 
    The gray region indicates a bistable regime, that we can access by following adiabatically the time-crystal solution as the interaction parameter $J$ is increased. 
    When starting from the ground state, the systems goes to the stationary phase before the end of the bistable region. 
    The other parameters are $\Omega=2\kappa$ and $J_z=\kappa$.
    (b)-(c) Stored energy (in units of $\omega_{\rm at}$) and efficiency as function of time for a time-crystal solution (red solid line) and a stationary one (blue dashed line). In both cases, the initial conditions were $\vec{m}=[0,0,-1/\sqrt{2}]$ and for the time-crystal solution we set $J=3.4\kappa$ while for stationary we use $J=3.41\kappa$. The vertical black lines 
    show that the second peak of efficiency correspond to the maximum stored energy. The remaining parameters are: $\Omega=2J_z=2\kappa$ and $\omega_{\rm at}=\nu$.}
    \label{fig:batteries}
\end{figure}

For the setup considered in this section [cf.~Fig.~\ref{fig:Fig1}\bc{(b)}], the reference Hamiltonian is ${H_{\rm ref} = H_{\rm at}^{(1)} + H_{\rm at}^{(2)}}$ and we further consider all atoms initially in the ground state. In this case, we have $\Delta u_\mathrm{s}(t)\approx \mathcal{E}(t) \ge 0$. 
Fig.~\ref{fig:batteries}\bc{(a)} shows that, averaging over the charging time, the time-crystal phase can store large amounts of energy. 
Note that there is a bistable region, in gray, in which the stationary solution and the time-crystal solution coexist. Upon varying the parameter $J$, the system jumps from the time-crystal phase to the stationary one and the amount of stored energy drastically decreases. 
In Fig.~\ref{fig:batteries}\bc{(b)-(c)} we plot the stored  energy and the efficiency of the charging process as a function of time. 
Here, we compare the time-crystal phase with the stationary one. To this end, we consider two  values of the interaction strength $J$  which are extremely close one another, but with one giving rise to a time-crystal and the other to a stationary phase.
Essentially, one interaction strength is on the left and the other one on the right of the value at which the stored energy drops in Fig.~\ref{fig:batteries}\bc{(a)}.
We clearly see that  the time-crystal phase allows to store more energy than the stationary phase [cf.~Fig.~\ref{fig:batteries}\bc{(b)}]. 
These peaks could serve as optimal times for  energy extraction.
In Fig.~\ref{fig:batteries}\bc{(c)}, we also see that the charging efficiency, as a function of time, is typically higher in the time-crystal regime. 

\section{Charging by seeding crystallization in time}\label{sec:setup2}

\begin{figure}
    \centering
     \includegraphics[width=\linewidth,height=\textheight,keepaspectratio]{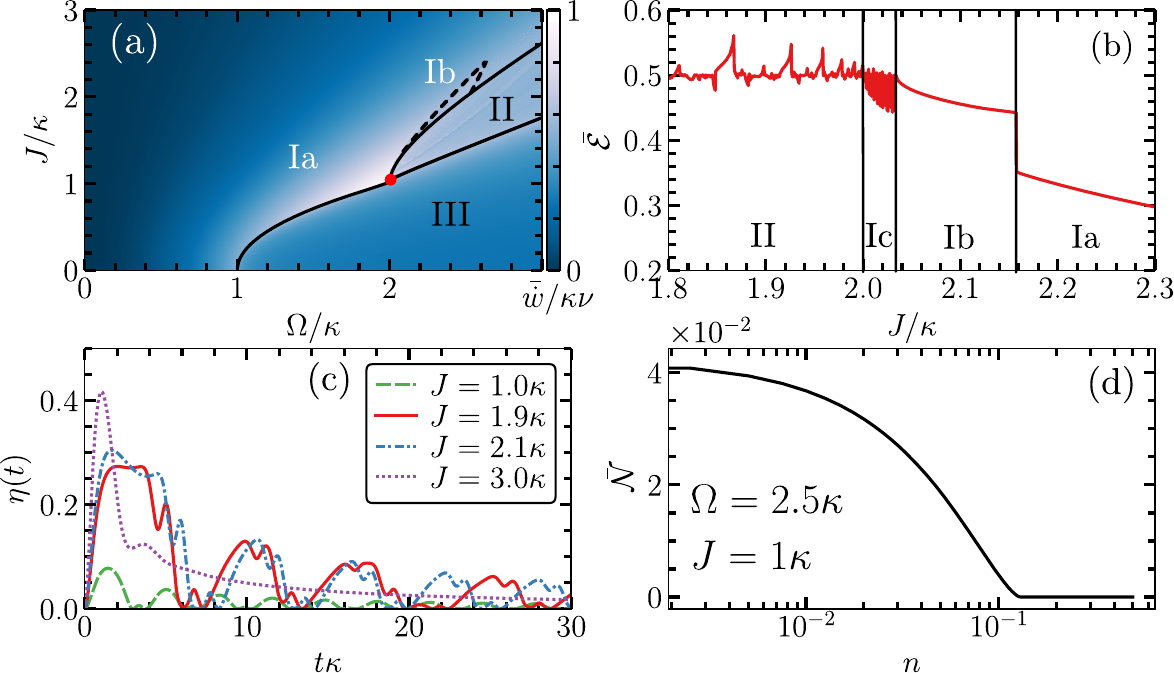}
    \caption{\textbf{Work input, efficiency, and entanglement.} (a) Time-integrated average work input as a function of $J$ and $\Omega$. We compute this phase-diagram by considering all atoms initially in the ground state for all points. We identify four regions: stationary (Ia), limit cycle (Ib), oscillatory regime between limit cycle and stationary (Ic), quasi-periodic oscillations (II) and limit cycle (III). (b) A section of the phase diagram (a) for the time-averaged stored energy in units of $\omega_{\rm at}$ as function of $J/\kappa$ for $\Omega=2.5\kappa$. In region II, we have the stored energy in the quasi-periodic regime. In Ic, the system oscillates between limit-cycles and stationary values for small variations of  $J/\kappa$. After this intermediate region, it goes to region Ib until it reaches  region Ia. Here we time-average trajectories with total time $t\kappa=10^3$. The stored energy is given  in units of $\omega_{\rm at}$. (c) The efficiency $\eta(t)$ as function of time, for four values of $J$, as indicated in the legend. Here $\Omega=2.5\kappa$. The blue dashed line correspond to a limit-cycle solution found in region I with initial state being the ground state of $H_{\rm at}$ (d) Time-averaged entanglement negativity between the charger and the battery as function of the occupation number ${n}_1={n}_2={n}$. We further set   $\omega_{\rm at}=\nu$.}
    \label{fig:seeding}
\end{figure}

We now investigate the energy storage and its efficiency in  a battery which is charged via seeding crystallization in time~\cite{PhysRevLett.128.080603}. 
In this setup, [cf.~Fig.~\ref{fig:Fig1}\bc{(c)}], we consider one atomic ensemble without external driving as the battery,  while the second atomic ensemble, referred to as the charger, is driven by a laser.  
For simplicity, we consider, apart from the laser driving, both atomic systems with the same parameters, and $\delta=J_z=0$.
In addition to that, we consider initial conditions with $m_x^{(2)} = m_y^{(1)} = 0 $, which leads to $m_x^{(2)}(t) = m_y^{(1)}(t)=0$ for all times. 
Given that the the norm of the vector $\vec{m}^{(j)}(t)=[m_x^{(j)}(t),m_y^{(j)}(t),m_z^{(j)}(t)]$ is conserved in each system, we can express the magnetizations as  
 \begin{eqnarray}\label{eq:Setup2_magnetizations}
    &m_x^{(1)}(t)=m_0 \sin[f^{(1)}(t)],\quad m_z^{(1)}(t)=m_0 \cos[f^{(1)}(t)] , \nonumber\\
   &m_y^{(2)}(t)=m_0 \sin[f^{(2)}(t)]  ,\quad m_z^{(2)}(t)=m_0 \cos[f^{(2)}(t)],
\end{eqnarray} 
where $m_0=-1/\sqrt{2}$.
By exploiting the aforementioned conditions, one can show that the dynamics of the coupled system is determined by 
\begin{eqnarray}\label{eq:setup2_mf}
    \dot{f}^{(1)}&= -J \sin(f^{(2)}) - \kappa \sin(f^{(1)}) ,\nonumber\\
    \dot{f}^{(2)} &= J \sin(f^{(1)}) - \kappa \sin(f^{(2)}) - \Omega .
\end{eqnarray}
This dynamical system features a rich phase diagram containing limit cycles, quasi-periodic oscillations, stationary solutions, as well as bistable regimes, as illustrated in Fig.~\ref{fig:seeding}\bc{(a)}. 
In the stationary regime, the stationary solutions of Eq.~\eqref{eq:setup2_mf} are given by 
\begin{eqnarray}
        m_z^{(1)}  =  \frac{\pm1}{\sqrt{2}}\sqrt{1 - \frac{(J\Omega)^2}{(J^2  +  \kappa^2)^2}}  ,\qquad 
        m_z^{(2)}  =   \frac{\pm1}{\sqrt{2}}\sqrt{1  -  \frac{(\kappa\Omega)^2}{(J^2  +  \kappa^2)^2}} ,
\end{eqnarray}
where only the negative $m_z^{(j)}$ is stable. 
Here, we observe two conditions that have to be simultaneously satisfied for the existence of stationary solutions, $\mathcal{C}_1: J^2 +  \kappa^2 \ge \kappa \Omega$ and $\mathcal{C}_2: J^2 +  \kappa^2 \ge J\Omega$.  
Using these conditions, we can find the critical interaction strength determining the transition between the stationary and non-stationary phases,  
\begin{equation}\label{eq:setup2_critical}
    J_\mathrm{c}  =  \max\limits_{ J\in \mathbb{R}}\left( J  =  \sqrt{\Omega \kappa  -  \kappa^2}, 2J  =   \Omega  +  \sqrt{\Omega^2  -  4\kappa^2}\right) .
\end{equation} 
We note the same behavior observed in the first setup, with strong interactions driving the system towards a  stationary phase. 

In Fig.~\ref{fig:seeding}\bc{(a)} we show the phase diagram for the time-averaged total work input in both battery and charger starting from the initial condition in which all atoms are in the ground state. 
Under these conditions, we can identify four distinct dynamical regions. 
We have region Ia featuring stationary solutions, region II displaying quasi-periodic solutions, region III characterized by one limit cycle solution, and region Ib characterized by limit-cycle phase when starting from all atoms down (a stationary solution is also possible when starting from a different initial condition).
From the condition in Eq.~\eqref{eq:setup2_critical}, we identify a tricritical point [red dot in Fig.~\ref{fig:seeding}\bc{(a)}], given by $J=\kappa$ and $\Omega =2\kappa$, between regions I, II, and II, as well as the curve separating the stationary phase from the other ones. 
We can interpret the emergence of quasi-periodicity as follows: 
in region I, both stationary conditions are satisfied, while in region III, the condition for non-stationarity $\mathcal{C}_2$ is not valid, regardless of the validity of $\mathcal{C}_1$.
For this region, the charger seeds time crystallization in the battery.
Conversely, in region II, only $\mathcal{C}_1$ is not satisfied.
This suggests that the competition between the battery preferring an oscillatory solution while the charger a stationary one gives rise to the quasi-periodic regime.
Moreover, in the stationary region, the total work input assumes the same form as the setup we considered in Sec.~\ref{sec:setup1}, i.e. ${\bar{\dot{w}} = \kappa \nu  \Omega^2/[J^2 + \kappa^2]}$.

For this setup [cf.~Fig.~\ref{fig:Fig1}\bc{(c)}] we consider that energy can only be stored in the atomic ensemble which is not directly driven by the laser, i.e.,  ${H_\mathrm{ref} = \omega_\mathrm{at} S_z^{(1)} / \sqrt{2}}$.
Recalling that $ H_\mathrm{s}  \approx  H_\mathrm{at}^{(1)} + H_\mathrm{at}^{(2)}$ due to our assumption  $\omega_\mathrm{at}\gg J, J_z, \Omega$ and considering all atoms initially in the ground state, the efficiency $\eta(t)$ is always between $0$ and $1$, since $\Delta u_s(t) \ge \mathcal{E}(t)$.
Here, we consider a cut in the phase diagram with a fixed laser driving and we vary the interaction, Fig.~\ref{fig:seeding}\bc{(b)}.
For these parameters, we evaluate the time-averaged stored energy in the process.
We observe that in the quasi-periodic phase the system can store, on average, more energy than in the other phases.  
Also, in region Ib the system  attains higher values of time-averaged stored energy. 
However, when the system parameters belong to region Ia, the amount of stored energy drops considerably. 
We note that region Ib coexists with region Ia and in region Ic, the system oscillates between limit-cycles and stationary solutions for small variations of $J/\kappa$ (see   \ref{ap:setup_2} for more details on the presence of multistability in this parameter regime).

In Fig.~\ref{fig:seeding}\bc{(c)} we plot the efficiency $\eta(t)$ for fixed driving and varying coupling, showing that it reaches a maximum value during the transient phase of the stationary solution. After this, the efficiency assumes higher values for the quasi-periodic and limit cycle solutions. 
We interpret this result as follows.
A small $J$ limits the flux of energy from the charger to the battery, which makes the limit cycle phase in region III not efficient in storing energy, since the battery oscillates close to its ground state. 
On the other hand, in the region of quasi-periodicity and for limit cycles solutions in region Ib, the internal energy in the battery system oscillates around half of its maximum value and with high amplitude, thereby achieving large values of both $\mathcal{E}(t)$ and $\eta(t)$. 

In addition to the stored energy, we investigate correlation properties in this setup. We observe a finite entanglement negativity in the time-crystal phase, which vanishes as the temperature increases, as shown in  Fig.~\ref{fig:seeding}\bc{(d)}. 
To further understand this behaviour in the limit-cycle phase, we look at the Hamiltonian of the fluctuation operators 
\begin{equation}\label{eq:Hamiltonian_fluc_setup2}
    H_\mathrm{f}(t) = J\sqrt{2}[m_z^{(2)}(t)x_1x_2 + m_z^{(1)}(t)p_1p_2] .
\end{equation}
This Hamiltonian contains two time-dependent terms, that are related to two-mode squeezing and excitation exchange, and allow for entanglement during limit cycle phases~\cite{PhysRevA.108.062216}. 
The jump operators of system $j$, have the same form of the one reported for the other setup. 
The persistent entanglement may be further explored, for instance, for metrological  applications~\cite{Montenegro2023,Pavlov_2023,MA201189} (see   \ref{ap:fluctuations} for furthers details).

\section{Conclusions}

We have fully characterized of the non-linear dynamics and thermodynamics of two coupled boundary time crystals.
Using this characterization, we have investigated the coupled systems in the context of energy storage.
Our thermodynamic analysis allowed us to assess the efficiency of driven-dissipative quantum batteries operating in both stationary and oscillatory phases.
We have explored energy storage within two different setups and observed distinct features. 
In the first setup, where both atomic ensembles are regarded as the batteries and the charging is provided by external lasers following a parallel charging process, we have found that the time-crystal phase surpasses the stationary phase in  both the amount of stored energy and efficiency. 
In addition to that, we have shown that in this setup the batteries develop quantum correlations in both phases, but with non-zero entanglement only within the stationary phase.
In the second setup, where one atomic ensemble transfers energy to the other one, we have found that the dynamics is described by a set of  non-linear coupled phase equations. 
These equations provide the critical lines, separating the stationary solutions from the limit cycle and the  quasi-periodic solutions.
Regarding this setup, our analysis suggests that the optimal time for energy storage occurs during a transient phase toward the stationary state.
However, we also observe that when the system operates in oscillatory phases, it demonstrates high  efficiency in storing energy for longer periods of time.
Furthermore, we show that the battery and the charger can remain entangled during the limit cycle phase. 
%
Our findings suggest that coupled boundary time crystals can be used as quantum batteries that operate in both oscillatory and stationary phases.

The codes used to produce the
numerical results of this paper are available on Github~\cite{Github}.

\ack
We are grateful to Farokh Mivehvar and Parvinder Solanki for useful discussions.
We acknowledge funding from the Deutsche Forschungsgemeinschaft (DFG, German Research Foundation) under Project No. 435696605 and through the Research Units FOR 5413/1, Grant No. 465199066 and FOR 5522/1, Grant No. 499180199. This project has also received funding from the European Union’s Horizon Europe research and innovation program under Grant Agreement No. 101046968 (BRISQ). F.C.~is indebted to the Baden-W\"urttemberg Stiftung for the financial support of this research project by the Eliteprogramme for Postdocs. A.C. is grateful for financing from the Deutsche Forschungsgemeinschaft (DFG, German Research Foundation) through
the Walter Benjamin programme, Grant No. 519847240.
G.D.C. acknowledges support from the UK EPSRC through Grant No. EP/S02994X/1.
This work was funded by the QuantERA II Programme (project CoQuaDis, DFG Grant No. 532763411) that has received funding from the EU H2020 research and innovation programme under GA No. 101017733.
\appendix

\section*{Appendix}

\section{Thermodynamics of collision models}\label{ap:thermodynamics}

In this Appendix, we provide additional details on the derivation of a consistent thermodynamic description for the coupled boundary time crystal system. In order to do so, we consider the dynamics of the system composed by the coupled atomic ensembles together with the ancillae, defining the collision model.  Our derivation follows closely the ones presented in Refs.~\cite{de2018reconciliation,carollo2023quantum}.

The dynamics of the atomic systems interacting with the ancillae is governed by the following Hamiltonian in the laboratory frame
\begin{eqnarray}
    H_k(t) &= \sum_{j=1,2}[ H_\mathrm{at}^{(j)} + H^{(j)}_{\mathrm{e}_k} + H^{(j)}_{\mathrm{se}_k}(t)] +H_\mathrm{int} \\
    &  + \sum\limits_{j=1,2}\frac{\Omega_j}{\sqrt{2}}(S_-^{(j)}e^{i\omega_\mathrm{las}t} + S_+^{(j)}e^{-i\omega_\mathrm{las}t}) ,
\end{eqnarray}
where we consider both lasers with the same frequency and the Hamiltonians for the atomic ensembles and $k$-th ancilla are defined in the main text. 
In this way, we now evaluate the quantities related to the first law of thermodynamics. 
To this end,  we consider the variation of the density matrix after a single collision between the system and an ancilla 
\begin{equation}
    \Delta \rho_{{\rm se}_k} = V_k\left(\rho_{\rm s} \otimes \rho_{{\rm e}_k} \right)V_k^\dag - \rho_{{\rm s}}\otimes \rho_{{\rm e}_k} ,
\end{equation}
where $\rho_{{\rm e}_k}=\rho_{{\rm e}_k}^{(1)} \otimes \rho_{{\rm e}_k}^{(2)}$  and 
\begin{equation}\label{eq:App_collision}
    V_k = \mathcal{T}e^{-i\int_t^{t+\delta t} H_k(t)\mathrm{d}t^\prime} ,
\end{equation}
where $\mathcal{T}$ is the time-ordering operator. 

The dissipated heat to the environment $j$ is given by the energy absorbed by the $k$-th ancilla during the collision with the system
\begin{equation}
    \Delta Q^{(j)}_k = - \lim\limits_{\delta t\rightarrow 0}\frac{1}{\delta t} {\rm Tr}_k\{\nu  (a_k^{(j)})^\dag a^{(j)}_k\Delta \rho_{{\rm se}_k}\} .
\end{equation}
In addition to that, the change in the internal energy of the system reads  
\begin{equation*}
   \Delta U_k = \lim\limits_{\delta t\rightarrow 0} \frac{1}{\delta t}{\rm Tr}_k\{ H_{\rm s}\Delta\rho_{{\rm se}_k} +  \Delta H_{{\rm s}}\rho_{{\rm s}}\otimes \rho_{{\rm e}_k}\}
\end{equation*}
where the second term accounts for the change in the Hamiltonian due the time-dependency.
In order to compute the quantities above, we can go to the Heisenberg picture
\begin{equation*}
    \Delta \langle \mathcal{O} \rangle = \langle V_k^\dag \mathcal{O} V_k - \mathcal{O}\rangle .
\end{equation*}
For calculating the above expression, we expand the unitary operator in Eq.~\eqref{eq:App_collision} up to the second order in $\delta t$, finding
\begin{equation*}
    \Delta \langle \mathcal{O} \rangle = i\delta t \langle [H_{k}, \mathcal{O}] \rangle - \frac{\delta t^2}{2} \langle  [H_{k}, [H_{k}, \mathcal{O}]] \rangle .
\end{equation*}
We use the limit $\delta t \ll \omega_{\rm las}^{-1}$, such that the time-dependent laser term remains approximately constant during one collision. 
Using that the ancillary system is in a product and thermal state, the dynamics of the Hamiltonians simplifies to 
\begin{eqnarray*}
    \Delta \langle {H}_{\rm s} \rangle &= - \frac{\delta t^2}{2} \langle  [H_{{\rm se}_k}, [H_{{\rm se}_k}, H_{\rm s}]] \rangle + \delta t\langle \dot{H}_{\rm s}\rangle,\\
    \Delta \langle {H}_{e_k}^{(j)} \rangle &= - \frac{\delta t^2}{2} \langle  [H_{{\rm se}_k}^{(j)}, [H_{{\rm se}_k}^{(j)}, H_{{\rm e}_k}^{(j)}]] \rangle .
\end{eqnarray*}
The quantity $\langle \dot{H}_{\rm s}\rangle$ is referred as an external work input from the lasers. In order to perform the calculations, it is more convenient to move to the frame rotating with the laser. This can be achieved through the unitary transform for the $k$-th collision
\begin{equation*}
    \mathcal{V}_k = e^{ -i\omega_{\rm las} t\left[ (a^{(1)}_k)^{\dag} a^{(1)}_k + (a^{(2)}_k)^\dag a^{(2)}_k + \frac{S_z^{(1)}}{\sqrt{2}} +  \frac{S_z^{(2)}}{\sqrt{2}}\right]} ,
\end{equation*}
such that $\tilde{\rho}_{{\rm se}_{k}} = \mathcal{V}_k\rho_{{\rm se}_k} \mathcal{V}_k^\dag$.
With this, we find the Hamiltonian in the rotating frame
\begin{eqnarray}
    H_{\rm s}^{\rm rot} &=  H_{\rm int} + H^{(1)}_{{\rm se}_k} + H^{(2)}_{{\rm se}_k}  \\
    & +\frac{\Omega_{1}}{\sqrt{2}} S_x^{(1)} +  \frac{\Omega_{2}}{\sqrt{2}} S_x^{(2)} + \frac{\delta}{\sqrt{2}} S_z^{(1)} +  \frac{\delta}{\sqrt{2}} S_z^{(2)}  \\
    & + \tilde{\delta} (a^{(1)}_k)^\dag a_k^{(1)} +  \tilde{\delta} (a_k^{(2)})^\dag a^{(2)}_k ,
\end{eqnarray}
where $\tilde{\delta} = \nu  - \omega_{\rm las}$.
In the first line, the Hamiltonians remain unchanged, in the second and third lines we have the Hamiltonians of the atomic systems and of the ancillae  in the rotating frame, respectively. 
We remark that when performing the collision dynamics within this Hamiltonian using $\delta t\rightarrow 0$, we find the Lindblad equation reported in the main text, Eq.~\eqref{eq:Lindblad}.

Here we provide details about the calculation of the dissipated heat. 
We start with the dissipated heat in the laboratory frame and we use identities $\mathcal{V}_k^\dag\mathcal{V}_k = \mathcal{V}_k\mathcal{V}_k^\dag = I$ and the cyclic property of the trace, such that we can use the state evaluated in the rotating frame to compute the observables of interest in the laboratory frame. 
In this way, we obtain the heat dissipated to the $j$-th  environment 
\begin{eqnarray}
    \Delta Q^{(j)} &=   -\Delta \langle {H}^{(j)}_{{\rm e}_k} \rangle \nonumber \\
    &=-\frac{\delta t \kappa \nu }{2N} {\rm Tr}\left\{ \left[S_+^{(j)}S_-^{(j)} + S_-^{(j)}S_+^{(j)} \right.\right.\\
    &\left.\left.+ 2\sqrt{2}S_z\left(2(a_k^{(j)})^\dag a_k^{(j)} + 1\right)\right] \tilde{\rho}_{{\rm se}_{k}}\right\}\nonumber .
\end{eqnarray}
We note that when considering a mesoscopic scale, $1 \ll N < \infty$, we can approximate the dissipated heat by the $j$-th system to~\cite{carollo2023quantum} 
\begin{equation}\label{eq:Large_N_heat}
     \dot{Q}^{(j)}(t) \approx  N \dot{Q}^{(j)}_{\rm mf}(t) + \dot{Q}_{\rm sub}^{(j)}(t) 
\end{equation} 
where $\dot{Q}^{(j)}_\mathrm{mf}(t) = -\kappa \nu [ (m_x^{(j)})^2 + (m_y^{(j)})^2]$ and 
\begin{equation}
    \dot{Q}_{\rm sub}^{(j)}=-\kappa \nu \left[ G_{x_jx_j}^N + G_{y_jy_j}^N + \sqrt{2} m_z^{(j)}(2n_j + 1) \right] ,
\end{equation}
where total dissipated heat is given by an extensive contribution from the mean-field, $\dot{Q}_{\rm mf}^{(j)}$, and a intensive contribution that accounts for the fluctuations effects, $\dot{Q}_{\rm sub}^{(j)}$. 
The matrix $G$ is defined in the   \ref{ap:fluctuations}.

Moreover, given that the collision model reproduces the Lindblad equation reported in the main text, the variation of the internal energy can be calculated through the equation
\begin{equation}
    \dot{U} = \sum\limits_{j,\alpha}\langle \mathcal{D}_{j,\alpha}^*[H_\mathrm{s}] \rangle +  \dot{W}_{\rm las}^{(1)}  +  \dot{W}_{\rm las}^{(2)},
\end{equation}
where $ \mathcal{D}_{j,\alpha}^*$ is the dual generator of the dissipative part of the dynamical generator, given by 
\begin{equation}
    D_{j,\alpha}^*[X] = \sum\limits_{j,\alpha} \left[  (L_\alpha^{(j)})^\dag X L_\alpha^{(j)}  -  \frac{1}{2}\left\{(L_\alpha^{(j)})^\dag L_\alpha^{(j)}, X \right\} \right] ,
\end{equation}
where $X$ is an arbitrary operator.
Here we note that the unitary $\mathcal{V}_k$ leaves the dissipative part unchanged. 
Finally, the work input from the lasers is calculated by considering the time-derivative of the Hamiltonian of the laser
\begin{equation}
    \dot{W}_{\rm las}^{(j)} = \frac{\Omega_j}{\sqrt{2}}\Tr\left[ \frac{{\rm d}}{{\rm d}t}(S_-^{(j)}e^{-i\omega_{\rm las}t} + S_+^{(j)}e^{+i\omega_{\rm las}t})\rho_{\rm s}\otimes \rho_{{\rm e}_k}\right] . 
\end{equation}
Then, we remove the time-dependency by inserting the identities $\mathcal{V}$, such that we find  
\begin{equation}
    \dot{W}_{\rm las}^{(j)} =  -i\frac{\omega_{\rm las}\Omega_j}{\sqrt{2}}\Tr\left\{(S_-^{(j)}- S_+^{(j)}) (\rho_{{\rm s}}\otimes \rho_{{\rm e}_k})_{\rm rot}\right\} = -\omega_{\rm las}\Omega_j\sqrt{2}\langle S_y^{(j)} \rangle  .
\end{equation}

\subsection{The second law}

In this section, we are going to derive the second law of thermodynamics for the coupled boundary time crystals. This follows closely the derivations presented in Refs.~\cite{carollo2023quantum,RevModPhys.93.035008}. 

We start by defining the states, $\rho_{{\rm se}_k}^\prime \equiv V_k \left( \rho_{\rm s} \otimes \rho_{{\rm e}_k}\right) V_k^\dag $, $\rho_{\rm s}^\prime \equiv {\rm Tr}_{k}[\rho_{{\rm se}_k}]$, $\rho^\prime_{{\rm e}_k} \equiv {\rm Tr}_{\rm s}[\rho^\prime_{{\rm se}_k}]$,  $\rho^{(1)\prime}_{{\rm e}_k} \equiv {\rm Tr}_{2}[\rho^\prime_{{\rm e}_k}]$, and $\rho^{(2)\prime}_{{\rm e}_k} \equiv {\rm Tr}_{1}[\rho^\prime_{{\rm e}_k}]$.
Here we remark that for the derivation of the first law of thermodynamics, we look to the energy exchanged between the system and the ancillae system. 
For the second law, however, we consider the information loss when we trace-out the ancillae degrees of freedom. 
In this way, we can express the entropy production for this system as
\begin{equation}
    \Sigma = \mathcal{I}_{{{\rm e}_k}}({\rm s}:{\rm e}_k) + S(\rho_{{\rm e}_k}^\prime|| \rho_{{\rm e}_k}) \ge 0 ,
\end{equation}
where $\mathcal{I}_{{\rm e}_k}({\rm s}:{\rm e}_k) = S(\rho_{{\rm e}_k}^\prime) + S(\rho_{\rm s}^\prime) - S(\rho_{{\rm se}_k}^\prime)$ denotes the mutual information between the ancilla after the $j$-th collision and the system. $S(A||B)={\rm Tr}[A\ln A - A\ln B]$ is the is the relative entropy between the states $A$ and $B$.
Here, $S(A)=-{\rm Tr}[A\ln A]$ is the von Neumann entropy. 
The positivity of the entropy production is valid since both mutual information and relative entropy are positive.
Moreover, we can rewrite the entropy production above as
\begin{equation}
    \Sigma = \Delta S- \Phi_k ,
\end{equation}
where $\Phi_k=\Phi^{(1)}_k + \Phi_k^{(2)}$, is the entropy flux from the system to the environments during the $k$-th collision, with  \begin{equation}
    \Phi_k^{(j)}={\rm Tr}[(\rho_{{\rm e}_k}^{(j)\prime} - \rho_{{\rm e}_k}^{(j)})\ln \rho_{{\rm e}_k}^{(j)}] .
\end{equation}
$\Delta S$ is the variation of the von Neumann entropy of the system.
When dividing the above equation by $\delta t$ and taking the limit of $\delta t$ to zero, we go to the continuous limit, with $\dot{\Sigma} = \dot{S} - \dot{\Phi}$. 
In our model, we consider that the ancillary systems are initially in a thermal state, given by $\rho_{{\rm e}_k}^{(j)} = e^{-\beta_j H_{{\rm e}_k}^{(j)}}/Z_{{\rm e}_k}^{(j)}$, where $Z_{{\rm e}_k}^{(j)}$ is the partition function of the thermal ancilla $k$ interacting with the $j$-th environment.
In this scenario, the entropy flux  reduces to  
\begin{equation}
    \dot{\Phi}^{(j)}(t) = -\beta_j \lim\limits_{\delta t \rightarrow 0}\frac{1}{\delta t} {\rm Tr}\{(\rho_{{\rm e}_k}^{(j)\prime} - \rho_{{\rm e}_k}^{(j)}) \nu  (a_k^{(j)})^\dag a_k^{(j)}\} = \beta \dot{Q}^{(j)}(t) .
\end{equation}
This result shows that when the ancillae are in a thermal state, the entropy flux to the environment is proportional to the heat current flowing between the system and the bath~\cite{carollo2023quantum,RevModPhys.93.035008}. 
In the thermodynamic limit, the heat currents are always negative for our model. 

\section{Mean-field dynamics }\label{ap:meanfield}

In this appendix we write down the full system of equations describing the mean-field dynamics of the coupled boundary time crystals. This is given by
\begin{eqnarray*}\label{eq:mean-field}
    \dot{m}_{x}^{(1)} =& -J_z\sqrt{2}m_{y}^{(1)}m_{z}^{(2)}+J\sqrt{2}m_{y}^{(2)}m_{z}^{(1)} \\
    &- \delta m_{y}^{(1)} + \kappa \sqrt{2} m_{x}^{(1)}m_{z}^{(1)} ,\\
    \dot{m}_{y}^{(1)} =& -J\sqrt{2}m_{z}^{(1)}m_{x}^{(2)} + J_z\sqrt{2}m_{x}^{(1)}m_{z}^{(2)}\\
    &- \Omega_{1}m_{z}^{(1)} + \delta m_{x}^{(1)} +\kappa \sqrt{2}m_{y}^{(1)}m_{z}^{(1)} ,\\
    \dot{m}_{z}^{(1)} =& J\sqrt{2}\left(m_{y}^{(1)}m_{x}^{(2)} - m_{y}^{(2)}m_{x}^{(1)}\right) \\
    &+ \Omega_{1}m_{y}^{(1)} - \kappa\sqrt{2}((m_{x}^{(1)})^2 + (m_{y}^{(1)})^2) ,\\ 
    \dot{m}_{x}^{(2)} =& -J_z\sqrt{2}m_{y}^{(2)}m_{z}^{(1)}+J\sqrt{2}m_{y}^{(1)}m_{z}^{(2)} \\
    &- \delta m_{y}^{(2)} + \kappa \sqrt{2} m_{x}^{(2)}m_{z}^{(2)} ,\\
    \dot{m}_{y}^{(2)} =& -J\sqrt{2}m_{z}^{(2)}m_{x}^{(1)} + J_z\sqrt{2}m_{x}^{(2)}m_{z}^{(2)}\\
    &- \Omega_{2}m_{z}^{(2)} + \delta m_{x}^{(2)} +\kappa \sqrt{2}m_{y}^{(2)}m_{z}^{(2)} ,\\
    \dot{m}_{z}^{(2)} =& J\sqrt{2}\left(m_{y}^{(2)}m_{x}^{(1)} - m_{y}^{(1)}m_{x}^{(2)}\right) \\
    &+ \Omega_{2}m_{y}^{(2)} - \kappa\sqrt{2}((m_{x}^{(2)})^2 + (m_{y}^{(2)})^2).
\end{eqnarray*}
We observe that, as in Ref.~\cite{carollo2023quantum}, the mean-field equations do not depend on the temperature of the baths. Moreover, this systems of equations reduces to Eq.~(\ref{eq:Symmetric_equations}) and Eq.~(\ref{eq:setup2_mf}), when considering the parameter choices and initial conditions for setup one, Sec.~\ref{sec:setup1} or two, Sec.~\ref{sec:setup2}, respectively.

\section{Dynamics of the fluctuations}\label{ap:fluctuations}

In this Appendix, we provide the detailed calculations for the dynamics of the covariance matrix. For convenience, we make use of a different notation as in the rest of the text.  
In this way, we define the quantum fluctuation operators as 
\begin{equation}
    F_j^N = \frac{S_j - \langle S_j \rangle}{\sqrt{N}}  .
\end{equation}
Here, instead of two indexes, one referring to the atomic system and the other to the operator, we have just one.
Our new index ranges from $1$ to $6$ such that it represents the ordered set $\{(x, 1), (y,1), (z,1), (x,2), (y,2), (z,2)\}$, where we have the coordinate and the system, respectively. 
We note that we include the upper index $N$ indicating that the fluctuation operator corresponds to a finite system.
The dynamics of the quantum fluctuations is fully characterized by the covariance matrix~\cite{PhysRevA.79.052327, MA201189}
\begin{equation}    
    G_{jk} = \lim\limits_{N\rightarrow \infty}G_{jk}^N, \quad G_{jk}^N= \frac{1}{2}\langle \{F_j^N, F_k^N \} \rangle .
\end{equation}
Moreover, we denote the symplectic matrix as $s=s^{(1)}\oplus  s^{(2)}$, and the Levi-Civita tensor as $\epsilon =  \epsilon^{(1)} \oplus \epsilon^{(2)}$. 
We note that the dynamics of the symplectic matrix $s$ can be computed directly from the dynamics of the mean-field operators.
For evaluating the dynamics of the covariance matrix, we introduce the matrix $K$, defined as 
\begin{equation}
    K_{jk} = \lim\limits_{N\rightarrow \infty} K^N_{jk}, \quad  \quad K_{jk}^N = \langle F_j^NF_k^N\rangle .
\end{equation}
With this, we note that $G = (K + K^{\rm T})/2$, so we can look at the dynamics of $K$ to characterize $G$.
For this purpose, we calculate $ \dot{K}_{jk} = \lim\limits_{N\rightarrow \infty} \langle \mathcal{L}^*[F_j^N F_k^N]\rangle$, 
where $\mathcal{L}^*$ is the dual of the  Lindblad generator. The dual can be rewritten in a more convenient way by rewriting the raising and lowering operators in terms of the x and y components, becoming 
\begin{eqnarray}\label{eq:dual_Lindblad}
        \mathcal{L}^*[O] &= i[H_\mathrm{s}, O] + \sum\limits_{jk}\left( \frac{A_{jk}}{2N}[[S_j,O], S_k]  + i \frac{B_{jk}}{2N}\{[S_j,O], S_k\}\right).
\end{eqnarray}
Here we have defined the matrices  $A = A^{(1)} \oplus A^{(2)}$ and $B = B^{(1)} \oplus B^{(2)}$ with
\begin{equation*}
    A^{(j)} = \kappa(2n_j +1)\left(\begin{array}{ccc}
        1 & 0 & 0\\
        0 & 1 & 0\\ 
        0 & 0 & 0\\ 
    \end{array}\right) , \; B^{(j)}= \kappa\left(\begin{array}{ccc}
        0 & -1 & 0\\
        1 & 0 & 0\\ 
        0 & 0 & 0\\ 
    \end{array}\right) .
\end{equation*}
Additionally, we can express the Hamiltonian of the coupled system as 
\begin{equation*}
    H = H_{\rm M} + H_{\rm L}, \quad H_{\rm M} = \sum_{jk} \frac{M_{jk}}{N} S_jS_k, \quad H_{\rm L} =\sum_j \frac{h_{jj}}{\sqrt{2}} S_j ,
\end{equation*}
with $h = {\rm diag}(\Omega_1, 0, \delta , \Omega_2, 0, \delta )$ and 
\begin{equation}
     {M} = \frac{1}{2}\left(\begin{array}{cccccc}
        0 & 0 & 0 & J & 0 & 0\\
        0 & 0 & 0 & 0 & J & 0\\
        0 & 0 & 0 & 0 & 0 & J_z\\
        J & 0 & 0 & 0 & 0 & 0\\
        0 & J & 0 & 0 & 0 & 0\\
        0 & 0 & J_z & 0 & 0 & 0\\
    \end{array}\right) ,
\end{equation}
where the factor $1/2$ deals with an over count.

For calculating the fluctuation dynamics, we start with the Hamiltonian part. This can be splitted in two terms by using the commutation relation
\begin{equation*}
    [H, F_l^NF_m^N] = [H,F_l^N]F_m^N + F_l^N[H,F_m^N] .
\end{equation*}
For the first term after in the right-hand side of the equation, we have 
\begin{eqnarray*}
    i[H_{\rm M}, F_l^N] &=i\sum_{jk}M_{jk}\Biggl( F_j^N [F_k^N, F_l^N] +\frac{\langle S_j \rangle}{N}[S_k, F_l^N] + \\
    &+[F_j^N, F_l^N]F_k^N +[S_j, F_l^N]\frac{\langle S_k \rangle}{N}\Biggr) .
\end{eqnarray*}
In order to find this expression, we have just added and subtracted the expected values inside the commutators.
Moreover, the terms regarding the commutation between the fluctuation operators can be factorized together and written as 
\begin{equation*}
    \sum_{jk}M_{jk}\left( F_j^N [F_k^N, F_l^N] + [F_j^N, F_l^N]F_k^N \right) \approx -i\sum_{j}C_{lj} F_j .
\end{equation*}
Here, the approximation is exact when $N\rightarrow\infty$ and we define 
\begin{equation*}
    C = s\left( M^{\rm T} + M\right) .
\end{equation*}
The terms with the diagonal part in the subsystems read
\begin{eqnarray*}
    i\sum_{jk}M_{jk}  \left( \frac{\langle S_j \rangle}{N}[S_k, F_l^N]+   [S_j, F_l^N]\frac{\langle S_k \rangle}{N}\right)   =  -\sum_{w}  D^{\rm M}_{lw}  \frac{S_w}{\sqrt{N}} ,
\end{eqnarray*}
where $ D^{\rm M}_{lw} = - \sqrt{2}\sum_{jk} m_k \left( M_{jk} + M_{jk}^{\rm T} \right){\epsilon}_{klw} $.
In this way, we have that
\begin{equation*}
    \lim\limits_{N\rightarrow\infty} i\langle [H_{\rm M}, F_l^N]F_w^N] \rangle = \left[(D^{\rm M} + C)K\right]_{lw} .
\end{equation*}
In summary, the full contribution of the quadratic Hamiltonians for the dynamics of the  covariance reads 
\begin{equation*}
    \lim\limits_{N\rightarrow\infty}  i\langle [H_{\rm M}, F_l^NF_w^N]] \rangle = \left[\left(D^{\rm M}  + C\right) K  +  K \left(D^{\rm M}  +  C \right)^{\rm T}\right]_{lw}    .
\end{equation*}
The term involving the matrix $K$ on the left side can be easily derived by exploiting the symmetry of the matrices and their indexes. 

Moreover, the linear part of the  Hamiltonian, $H_{\rm L}$, has a similar structure of the above equation, 
\begin{eqnarray*}
        i[H_{\rm L}, F_l^N] = \sum_k D_{lk}^{\rm L} \frac{S_k}{\sqrt{N}} ,
\end{eqnarray*}
where $D_{lk}^L = -\sqrt{2} \sum_{j}h_j {\epsilon}_{jlk} $.
The expected value for this term  reads
\begin{equation*}
    \lim\limits_{N\rightarrow\infty} i\langle [H_{\rm L}, F_l^N]F_k^N] \rangle = \left[D^{\rm L}K\right]_{lk} ,
\end{equation*}
where we have used that the mean value of the fluctuation operators is zero. 
The full Hamiltonian dynamics for the matrix $K$ reads  
\begin{equation*}
    \lim\limits_{N\rightarrow\infty} i\langle [H_{\rm L}, F_l^NF_m^N]] \rangle =  \left[D^{\rm L}K + K \left(D^{\rm L}\right)^{\rm T}\right]_{lm} ,
\end{equation*}
where the second term on the right-hand side of the equation follows by using the antisymmetric property of $\epsilon_{jkl}$. 
With this we conclude the analysis of the Hamiltonian part. 

We now address the dissipative contribution of the Lindbladian on the fluctuation dynamics. We first consider the the term multiplied by the matrix ${A}$ in Eq.~\eqref{eq:dual_Lindblad}.
The  contribution of this term follows by evaluating  the commutator 
\begin{eqnarray*}
    \frac{1}{N}\langle [S_j, [F_l^NF_w^N, S_k]] \rangle &= \\
    -\langle [F_j^N, [F_k,F_l^N]]F_w^N &- F_l^N[F_j^N, [F_k,F_w^N]] \\
    - [F_j^N, F_l^N][F_k,F_w^N] 
    &- [F_k,F_l^N][F_j^N, F_w^N]\rangle .
\end{eqnarray*}
When we consider the thermodynamic limit, the terms in the second line become zero.
This follows due to a further scaling $1/\sqrt{N}$ that appears after evaluating the commutators. 
The third line can be rewritten by means of the symplectic matrix 
\begin{equation*}
     [F_k^N, F_l^N][F_j,F_w^N]   
    + [F_j,F_l^N][F_k^N, F_w^N] = -{s}_{kl}{s}_{jw}  -{s}_{jl}{s}_{kw} .
\end{equation*}
Then, the expected value of this term becomes
\begin{equation*}
   \lim\limits_{N\rightarrow\infty}  \sum\limits_{jk}  \frac{{A}_{jk}}{2N}\langle [S_j, [S_k,F_l^NF_w^N]] \rangle =  - ({s} {A}{s})_{lw} .
\end{equation*}
We note that the sign changes due to the odd permutation in the indexes of the symplectic matrix. 

We now move to the last term, the one involving the matrix $B$. 
To this end, we start by rewriting the anti-commutator as 
\begin{eqnarray*}
    \frac{1}{N}\langle \{S_j, [F_l^N F_m^N, S_k]\} \rangle &= -\langle \{F_j^N, [F_k^N,F_l^N F_m^N]   \}\rangle - 2\frac{\langle S_j\rangle}{N}[S_k,F_l^N F_m^N] .
\end{eqnarray*}
To find this expression we simply added and subtracted the expected value. 
We now focus on the first term after the equality sign, since the last term will be similar to the Hamiltonian contributions. 
In this way, we can expand the commutator as
\begin{eqnarray*}  
    \langle \{F_j^N, [F_k^N,F_l^N F_m^N]   \}\rangle &\approx i\left({s}_{kl}\langle \{F_j^N, F_m^N \}\rangle + {s}_{km}\langle \{F_j^N, F_l^N \}\rangle \right) .
\end{eqnarray*}
The approximation sign is valid in the thermodynamics limit, since in this limit we have $\langle S_j S_k\rangle \rightarrow \langle S_j \rangle \langle S_k\rangle $. 
Then, after applying the sum we find 
\begin{eqnarray*}
    \lim\limits_{N\rightarrow\infty} i  \sum\limits_{jk} \frac{{B}_{jk}}{2N} (-\langle \{F_j^N, [F_k^N,F_l^N F_m^N]   \}\rangle) =({s}{B} G + G {B}{s})_{lm} .
\end{eqnarray*}
Finally, we evaluate the last term
\begin{eqnarray*}
        i\sum\limits_{jk} \frac{{B}_{jk}}{N}\left(-\frac{\langle S_j\rangle }{N}\langle [S_k,F_l^N F_m^N]\rangle\right)=\\
        i\sum_w\sum\limits_{jk} \sqrt{2}\frac{{B}_{jk}}{N}\frac{\langle S_j\rangle }{N}\langle i{\epsilon}_{klw}F_w^N F_m^N + i{\epsilon}_{kmw}F_l^NF_w^N \rangle . 
\end{eqnarray*}
Therefore, we can write the above expression as
\begin{equation*}
    \lim\limits_{N\rightarrow \infty}i\sum\limits_{jk} {B}_{jk}\frac{\langle S_j\rangle }{N}\langle [S_k,F_l^N F_m^N]\rangle = \left[ D^B K + K (D^B)^{\rm T}\right]_{lm} ,
\end{equation*}
where $D_{lw}^{\rm B} = \sum_{jk}  \sqrt{2}{B}_{jk} m_j {\epsilon}_{klw} $.

From these calculations, we have the full dynamics of the covariance matrix. 
In order to write the dynamics of the fluctuation operators more compactly, we define $D \equiv D^{{{\rm L}}} + D^{{{\rm M}}} + D^{{{\rm B}}}$.
Then, the final expression for the matrix $K$ reads 
\begin{equation*}
    \dot{K} = (D + Q) K  + K(D + Q)^{\rm T} + {s}{B} G + G\left({s}{B}\right)^{\rm T} - {s} {A}{s} .
\end{equation*} 
With this, we have the dynamics of the covariance matrix, $G = (K + K^{\rm T})/2$, given by the Lyapunov equation
\begin{equation}
    \dot{G} =  P G + GP^{\rm T} -sAs ,
 \end{equation}
where $P = D+C+ {s}{B}$. 
Here, we note that the matrix $D$ is anti-symmetric and it gives the evolution of the mean-field operators, such that we have $\dot{\vec{m}}=D(\vec{m}) \vec{m}$. The other two matrices capture the part that can not be accessed by the mean-field equations. 

For convenience, we rewrite the matrices composing the dynamics of the fluctuation operators
\begin{eqnarray}\label{eq:Matrices_fluctuations}
    (D^{\rm L})_{lk} = -\sqrt{2}\sum_j h_j \epsilon_{jlk} , \nonumber\\
    (D^{\rm M})_{lw} = -\sqrt{2}\sum_{jk} m_k (M_{jk} + M_{jk}) \epsilon_{klw} ,\\
    (D^{\rm B})_{lw} = \sqrt{2}\sum_{jk} B_{jk} m_w  \epsilon_{klw} \nonumber,
\end{eqnarray}
and 
\begin{equation}
    C = s(M+M^{\rm T}) .
\end{equation}

\subsection{Covariance matrix in the rotating frame}

In this section, we analyze the covariance matrix in a frame that rotates in the opposite direction of the mean-field dynamics, such that the commutation relations of the fluctuations are time-independent.
Particularly, in this reference frame the fluctuations obey a bosonic algebra, such that we can use the results for Gaussian bosonic mode, to calculate the classical correlations, $\mathcal{J}$, the quantum discord, $\mathcal{D}$, and the entanglement negativity,  $\mathcal{N}$~\cite{PhysRevLett.105.020503, PhysRevLett.105.030501}.

We start by considering a time-dependent rotation matrix $R(t)$ that has the property to evolve the mean-field equations.
In this sense, the time-derivative of the rotation reads $\dot{R}(t) = D R(t)$, where $D$ is defined in Eq.~\eqref{eq:Matrices_fluctuations}.
Here $m(t)=[m_x^{(1)},m_y^{(1)},m_z^{(1)},m_x^{(2)},m_y^{(2)},m_z^{(2)}]$ is the vector of the mean-field variables of both ensembles. 
We now consider the rotated covariance matrix ${\rm d}_t (\bar{G}) = {\rm d}_t (R^{\rm T} G R)$, which also obeys a Lyapunov equation.
The symplectic matrix assumes the bosonic form
\begin{equation}
   (\sigma)^{(j)} = (R^{\rm T}(t) s R(t))^{(j)} = \left(\begin{array}{ccc}
    0 & 1 & 0 \\ 
    -1 & 0 & 0 \\ 
    0 & 0 & 0 \\ 
   \end{array}\right) .
\end{equation} 
We now can define the new bosonic quantum fluctuations as $\vec{X}= R^{\rm T} \vec{F} = (x_1, p_1, w_1, x_2, p_2, w_2)$, where $\vec{F}=(F_j)_{j=1}^6$ is the vector of the fluctuation operators. 
From the rotated covariance matrix, one can evaluate the correlations within the bosonic system, as shown in the Refs.~\cite{PhysRevLett.105.030501}.
For convenience, we review the procedure for computing these correlations in~\ref{app:QuantumInformation}. 

Furthermore, one does not need the analytical form of $R(t)$ to evaluate the correlations between the atomic systems. 
However, having this information allows us to derive an effective Lindblad generator for the fluctuation operators, as reported in the main text in Eqs.~\eqref{eq:Hamiltonian_fluc_setup1}-\eqref{eq:Hamiltonian_fluc_setup2}.
In general the rotation matrix  $R(t)$ can not be found analytically, since it requires  the analytical solution of the   mean-field dynamics.
However, one can find the expression of this matrix in some scenarios. 
For the first setup, described in Sec.~\ref{sec:setup1}, in the regime of $J=J_z$ and with the atoms initially in the ground state, the mean-field variables have the property that $m_x(t)$ is zero for all times, as reported in Refs.~\cite{PhysRevLett.121.035301,PhysRevA.105.L040202}. 
By using this information and that the magnetization is conserved in each atomic system, 
\begin{equation}
    (m_x^{(j)})^2(t) + (m_y^{(j)})^2(t)+ (m_z^{(j)})^2(t)= (m^{(j)})^2(0) ,
\end{equation}
we find $R(t)=R^{(1)}(t) \oplus R^{(2)}(t)$, with
\begin{equation}\label{eq:rotation_SC}
    R^{(j)}(t) = \sqrt{2}\left(\begin{array}{ccc}
        1/\sqrt{2} & 0 & 0 \\ 
        0 & m_z^{(j)}(t) & m_y^{(j)}(t) \\ 
         0 & -m_y^{(j)}(t) & m_z^{(j)}(t) \\ 
    \end{array} \right).
\end{equation}
Here, the rotation matrix is the same for each atomic system, $R^{(j)}(t)=R^{(2)}(t)=R^{(2)}(t)$, since in this regime, they exhibit  the same mean-field equations. 
The fluctuation operators in this frame assume the form 
\begin{eqnarray}
    x_j &= F_x^{(j)} ,\nonumber\\
    p_j &= \sqrt{2}(m_z^{(j)}F_y^{(j)} - m_y^{(j)}F_z^{(j)}) ,\\
    w_j &= \sqrt{2}(m_y^{(j)}F_y^{(j)} + m_z^{(j)}F_z^{(j)}) \nonumber.
\end{eqnarray}
Although we have the fluctuation operator $w_j$, it does not contribute to the dynamics of the correlation matrix in the bosonic frame, since it commutes with both $x_j$ and $p_j$. 

In the second setup, described in Sec.~\ref{sec:setup2}, we have an analogous property in the mean-field dynamics, wherein each system can be represented by a phase, as we write in Eq.~\eqref{eq:Setup2_magnetizations} of the main text.
Using this property, we find the form of the rotation matrix of setup two, $R(t)=R^{(1)}(t) \oplus R^{(2)}(t)$, where $R^{(1)}(t)$ is given by Eq.~\eqref{eq:rotation_SC}, and
\begin{equation}
        R^{(2)}(t) = \sqrt{2}\left(\begin{array}{ccc}
        m_z^{(2)}(t) & 0 & m_x^{(2)}(t) \\ 
        0 & 1/\sqrt{2} & 0 \\ 
       -m_x^{(2)}(t)  & 0 & m_z^{(2)}(t) 
    \end{array}\right).
\end{equation}

\subsection{Quantum information for Gaussian systems}\label{app:QuantumInformation}

We now review how one can calculate the quantum and classical correlations in Gaussian systems~\cite{PhysRevLett.105.030501}.
When we apply this rotation to the covariance matrix we find that it can be expressed as 
\begin{equation}
    2\bar{G} = 2RGR^{\rm T} =
     \left(\begin{array}{ccc}
    \alpha & \gamma \\
    \gamma^{\rm T} & \beta
\end{array}\right) ,
\end{equation}
where the coefficients $\alpha$, $\beta$, and  $\gamma $ are matrices representing the covariance between the new bosonic variables. 
Moreover, we define the symplectic invariants $c_\alpha = \det \alpha$,  $c_\beta = \det \beta$, $c_\gamma = \det \gamma$, and  $c_\delta = \det \delta$.
The covariance matrix will correspond to a physical state if $c_\alpha, c_\beta \ge 1$ and $\lambda_\pm \ge 1$, where $\lambda_\pm$ are the symplectic eigenvalues defined as $2\lambda_\pm^2 = \Delta \pm \sqrt{\Delta^2 - 4c_\delta}$, where $\Delta = c_\alpha + c_\beta + 2c_\gamma$. 
We further define the function
\begin{equation}
    f(x) = \left(\frac{x+1}{2}\right)\log \left(\frac{x+1}{2}\right) - \left(\frac{x-1}{2}\right)\log \left(\frac{x-1}{2}\right) .
\end{equation}
Using this equation, one can evaluate the von Neumann entropy of the system using that $S=\sum_{m=1}^N f(\lambda_m)$, where $\lambda_m$  are the symplectic eigenvalues. 
This can be interpreted in terms of the Williamson theorem~\cite{Williamson_theo}, which states that the symplectic eigenvalues of the covariance matrix represents a thermal state given as $    \rho_{\rm th } = \rho_{\rm th}^{(1)} \otimes \rho_{\rm th}^{(2)}$, where 
\begin{equation}
\rho_{\rm th}^{(j)} = \sum_{k=0}^\infty \frac{\bar{n}_j^k}{(\bar{n}_j+1)^{(k+1)}} |k\rangle \langle k| ,
\end{equation}
where $|k\rangle$ is a Fock state and $\bar{n}_j$ is a occupation number of the $j$-th mode and it relates to the respective symplectic eigenvalue of the $j$-th mode as $\bar{n}_j = (\lambda_j-1)/2$. 
In other words, the symplectic transformation removes the Gaussian unitary operations in the state, bringing the system to a locally thermal state. 

Furthermore, by using these quantities we can compute the one-way quantum discord and the one-way classical correlation, respectively, 
\begin{eqnarray}
    \mathcal{D} &= f(\sqrt{c_\beta}) - f(\lambda_+) - f(\lambda_-) + f(\sqrt{E_{\rm min}}) \\
    \mathcal{J} &= f(\sqrt{c_\alpha}) - f(\sqrt{E_{\rm min}}) ,
\end{eqnarray}
where 
\begin{eqnarray*}
        E_{\rm min} &= \frac{2c_\gamma^2 + (c_\beta - 1)(c_\delta - c_\alpha)}{(c_\beta - 1)^2} +\frac{ 2|c_\gamma|\sqrt{c_\gamma^2 + (c_\beta - 1)(c_\delta - c_\alpha)}}{(c_\beta - 1)^2} 
\end{eqnarray*}
if $ (c_\delta - c_\alpha c_\beta)^2 \leq (1 + c_\beta)(c_\gamma^2)(c_\alpha + c_\delta)$ and 
\begin{eqnarray*}
        E_{\rm min} &=       \frac{c_\alpha c_\beta - c_\gamma^2 + c_\delta}{2c_\beta}- \frac{\sqrt{c_\gamma^4 + (c_\delta - c_\alpha c_\beta)^2 - 2c_\gamma^2(c_\alpha c_\beta + c_\delta)}}{2c_\beta} ,
\end{eqnarray*}
otherwise. 

The last quantity we consider is the logarithmic negativity, which is a witness of the amount of bipartite entanglement between the two bosonic systems. 
This quantity is given by 
\begin{equation}
    \mathcal{N} = \max(0, -\ln \tilde{\lambda}_-) ,
\end{equation}
where $\tilde{\lambda}_-$ is the smallest symplectic eigenvalue of the partially transposed covariance matrix, obtained by replacing $c_\gamma$ to $-c_\gamma$~\cite{PhysRevLett.105.030501}.

\section{Additional information for setup one}

In this section we provide further information about the results concerning the setup one for the mean-field and fluctuation dynamics.

\begin{figure}[t]
    \centering
    \includegraphics[width=\linewidth,height=\textheight,keepaspectratio]{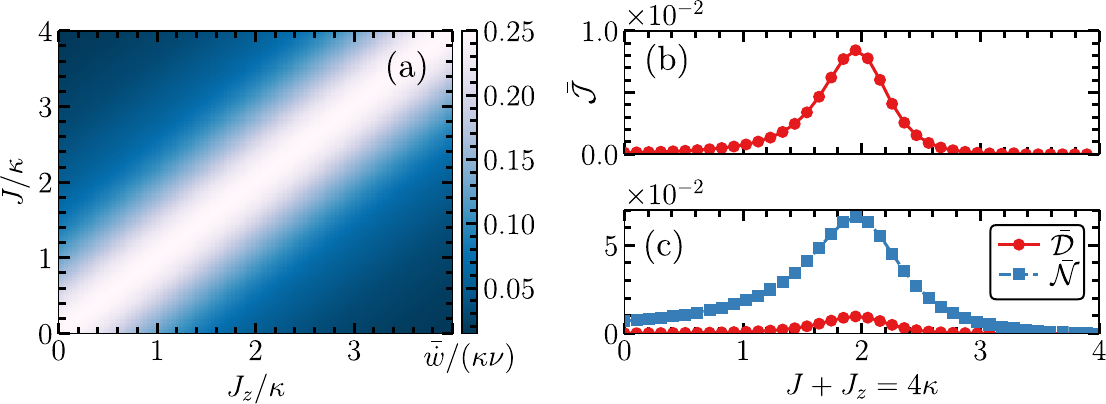}
    \caption{\textbf{Time-averaged work input and quantum correlations for the stationary phase.} (a) Time-averaged work input for varying $J_z$ and $J$, for the stationary phase  $\Omega_1=\Omega_2=\kappa/2$, respectively. Here, we start with $J=0$ and $J_z=4\kappa$ and we increase $J$ and decrease $J_z$. (b)-(c) Time-averaged classical and quantum correlations, respectively, as function of $J$, such that $J_z + J = 4\kappa$. The mean-field quantities were evolved from $t=0$ to $t\kappa=10^3$ and only their second half were integrated. The correlations were computed considering time intervals  $t \in [0, 200]$, and the integration was done over the second half of the interval. The remaining parameters are $\delta=0$, $n_1 = n_2 = 0$ and $\kappa =  2\Omega_1 = 2\Omega_2$. The initial state is $\vec{m}=[0,0,-1/\sqrt{2}]$. }
    \label{fig:Thermodynamics_SP}
\end{figure}

We start by analysing the time-averaged work input and correlations for the system in the stationary phase emerging when  $\Omega^2 < \kappa^2 + (J-J_z)^2$. In   Fig.~\ref{fig:Thermodynamics_SP}\bc{(a)} we show the total time-averaged work input in the system. This can be understood from the analytical solutions of the stationary phase from which we obtain an analytical expression for this quantity $\bar{\dot{w}} = \kappa \nu \Omega^2/[(J-J_z)^2 + \kappa^2]$. In Fig.~\ref{fig:Thermodynamics_SP}\bc{(b)-(c)} we show the behaviour of the correlations. 
In contrast to the results for the time-crystal phase, we observe that both correlations are maximized when $J=J_z$. 

\begin{figure}[t]
    \centering
    \includegraphics[width=\linewidth,height=\textheight,keepaspectratio]{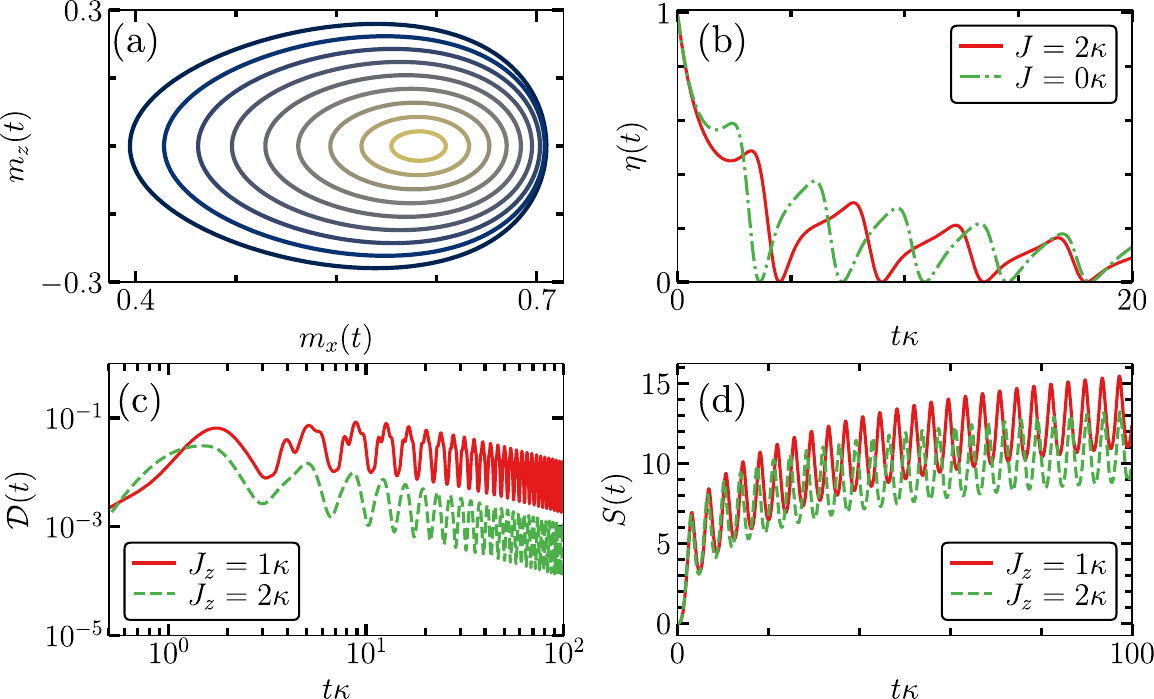}
    \caption{\textbf{Continuous family of oscillations, efficiency, quantum discord, and entropy.}  (a) Time-crystal solutions accessed with different initial conditions, $m_z(0) =0$, $m_x(0) = \sqrt{1/2}\cos(\phi)$ and $m_y(0) = \sqrt{1/2}\sin(\phi)$. Here we vary $\phi$ from $0$ to $\pi/6$ in ten intervals. The amplitude of the solutions decrease as $\phi$ increases, being zero when $\phi=\pi/6$. Also, we consider $J=\Omega=2J_z=2\kappa$.  In (b), we plot the efficiency when the system starts in the ground state for different $J$, as indicated in the legend, for $\Omega=2\kappa$ and $J_z=0$.
    Moreover, in (c)-(d) we plot the quantum discord and the entropy as function of time, respectively. In these plots, we consider $J=\Omega=2\kappa$ and $\omega_{\rm at}=\nu$.}
    \label{fig:Setup1_appendix}
\end{figure}

We now address in more detail the continuous family of oscillatory solutions that emerge in the time-crystal phase  due to the presence of the conserved quantity $\Gamma$ [see Eq.~\eqref{eq:Symmetric_equations}]. 
To this end, we exemplify in Fig.~\ref{fig:Setup1_appendix}\bc{(a)} some of the time-crystal solutions that can be accessed through different initial conditions.
The figure shows that the amplitude in the $z$ direction decreases until it reaches zero, thereby becoming a stationary point in the time-crystal region, for when $m_z(0)=0$, $m_x(0)=\cos(\pi/6)$, and $m_y(0)=\sin(\pi/6)$. 
Moreover, in Fig.~\ref{fig:Setup1_appendix}\bc{(b)} we show a comparison between two time-crystal solutions for different $J$, showing that the interaction strength does not change much the efficiency. 
This fact is supported by the results shown in  Fig.~\ref{fig:batteries} in the main text, where we observe that the time-averaged stored energy does not depend on $J$ inside the time-crystal region. 
Moreover, in Fig.~\ref{fig:Setup1_appendix}\bc{(c)-(d)}, we show the discord and the entropy for two scenarios, respectively. 
In the first one, given in red solid lines, we have $J\neq J_z$, and $J=J_z$ is given by the green dashed line. 
This shows the asymptotic behavior of both quantities, where the quantum correlations increase in the beginning of the dynamics, and then they decrease in time with a power law.
On the other hand, the entropy of both systems increases.
This behavior of the entropy is analogous to the one observed in the single boundary time-crystal, as reported in the Refs.~\cite{PhysRevA.91.051601,PhysRevA.105.L040202} 
Here, we note that when the interactions are different, the system entropy grows with a higher rate.

\section{Additional information for setup two}\label{ap:setup_2}

\begin{figure}[t]
    \centering
    \includegraphics[width=\linewidth,height=\textheight,keepaspectratio]{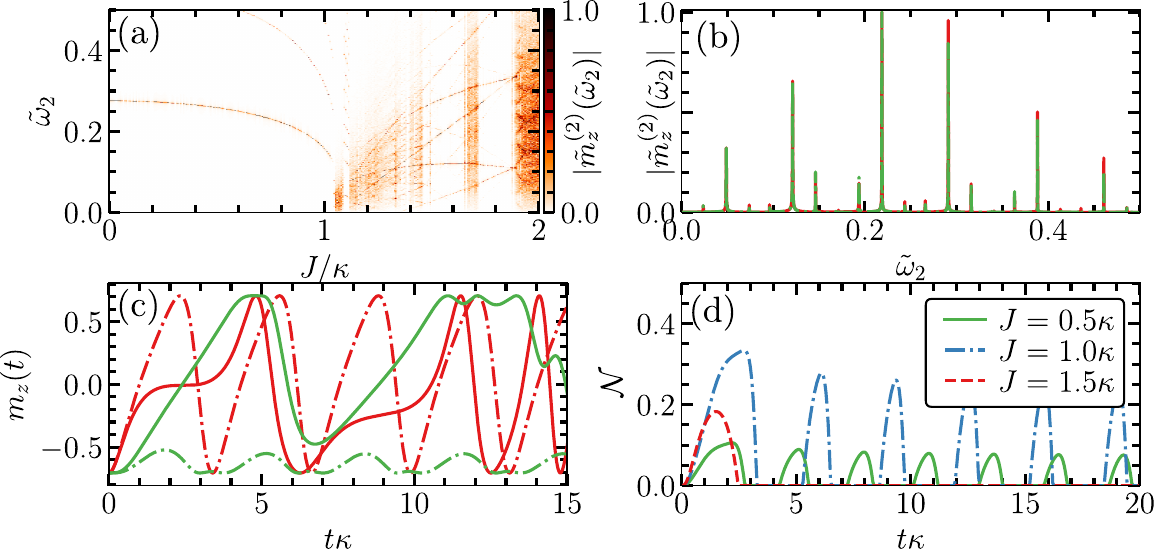}
    \caption{ \textbf{Fourier spectrum and trajectories of magnetization and entanglement.} (a) Density plot of the Fourier spectrum of $|\tilde{m}_z^{(2)}(\tilde{\omega}_2)|$ as function of $J/\kappa$ for the configuration in the setup two. In (b), we plot a section of the spectrum in (a) with $J=1.6\kappa$ for the battery (red solid line) and the charger (green dot dashed line). In (c) we compared the behaviour of $m_z(t)$ for the battery (green lines) and the charger (red lines). The solid lines refer to a solution with $J=1.5\kappa$, while the dot dashed lines to a solution with $J=\kappa$, for $\Omega=2.5 \kappa$.
    (d) Negativity for different values of $J$, as indicated in the legend. Here, $\Omega=2.5\kappa$. }
    \label{fig:Setup2_appendix}
\end{figure}

In this section, we provide further information on the mean-field and quantum fluctuations dynamics concerning the setup two. 

We start by investigating the Fourier spectrum of the magnetization in the $z$-direction for a fixed value of Rabi frequency and different values of $J$, as shown by Fig.~\ref{fig:Setup2_appendix}\bc{(a)}.
In this figure, we observe the difference between the limit-cycle phase, displaying few well-resolved spectral peaks, and the quasi-periodic phase, in which the spectrum displays many nearby peaks. 
Also, we observe that as $J$ varies, the Fourier spectrum in the quasi-periodic region varies in a non-trivial way, exhibiting sudden changes in the density of peaks. 
In  Fig.~\ref{fig:Setup2_appendix}\bc{(b)} we show the Fourier spectrum for one case in the quasi-periodic phase. 
Notice that both atomic ensembles always display the same Fourier peaks, since their dynamics is synchronized. 
We remark that this behavior is not chaotic given that, due to the existence of conserved quantities, the effective manifold of dynamical equations has effectively only two dimensions. 

Moreover, in Fig.~\ref{fig:Setup2_appendix}\bc{(c)} we plot the mean-field magnetisation $m_z(t)$ for the charger and for the battery. 
Our results show that for a small $J$, the battery undergoes a limit cycle having an average amplitude close to the ground state, thereby not being able to store much energy. 
However, when the system operates in the quasi-periodic phase, it oscillates between the minimum and the maximum of energy stored. 
Furthermore, in Fig.~\ref{fig:Setup2_appendix}\bc{(d)}, we show trajectories for the entanglement negativity for different $J$.
We see that the entanglement increases from $J=0.5\kappa$  to $J=\kappa$, but it becomes zero in the quasi-periodic phase. 

\begin{figure}
    \centering
    \includegraphics[width=\linewidth,height=\textheight,keepaspectratio]{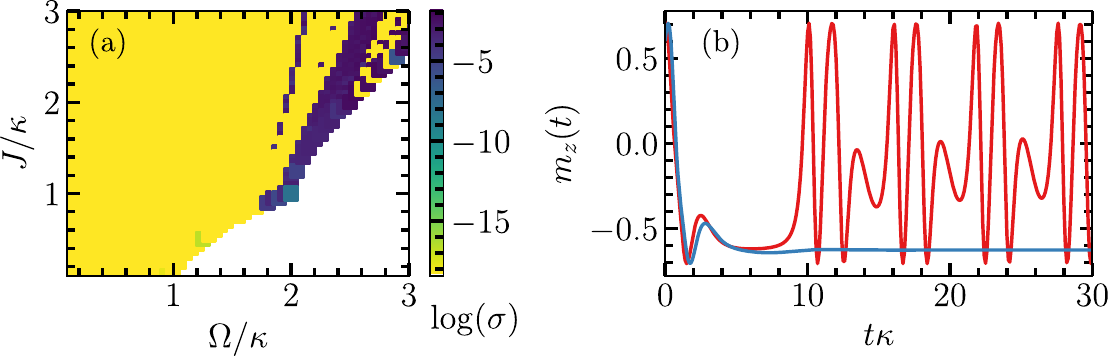}
    \caption{\textbf{Multistability} (a) Multistability in the stationary phase. In order to obtain this color plot, for each parameter value, we have time evolved 20 different trajectories of $m_z^{(2)}$ with random initial conditions. For each of these trajectories, we compute the time-average of $m_z^{(2)}$ between times $t\kappa=180$ and $t\kappa=200$. The color scale corresponds to the variance over trajectories of this time-averaged magnetization, $\sigma$. If there is a unique stationary solution, this variance is zero, otherwise there are multiple solutions. In (b) two trajectories of the magnetization of system two,  obtained from two different initial conditions. For the time-crystal solution we use $m_x^{(1)}(0)=m_y^{(2)}(0)=0$ and $m_z^{(1)}(0)=m_z^{(2)}(0)=1/\sqrt{2}$. For the stationary solution, we set $m_x^{(1)}(0)=0$, $m_z^{(1)}(0)=1/\sqrt{2}$, $m_x^{(2)}(0)=\sin(\pi/2)/\sqrt{2}$, and $m_z^{(2)}(0)=\cos(\pi/2)\sqrt{2}$ .  Here, we consider $\Omega=2.5\kappa$ and $J=2.1\kappa$.}
    \label{fig:Setup2_appendix_bistability}
\end{figure}

Finally, we investigate the presence of multistability for the parameters considered in Fig. \ref{fig:seeding}{\bc (a)}, when allowing for different initial conditions. 
In Fig.~\ref{fig:Setup2_appendix_bistability}{\bc (a)}, we show that within the stationary region, there are regions where oscillatory solutions are also possible.
Dark regions correspond to regions where there is coexistence between oscillatory and stationary solutions.
In Fig.~\ref{fig:Setup2_appendix_bistability}{\bc{(b)}} we exemplify the presence of multistability by considering two initial conditions for the same parameter choice, and observing that in one case the system relaxes to a stationary point (blue line) while in the other it goes to a limit-cycle solution (red line).

\section*{References}
\bibliographystyle{iopart-num}
\bibliography{references}

\end{document}